\documentclass[lettersize,journal]{IEEEtran}
\usepackage{amsmath,amsfonts}
\usepackage{algorithmic}
\usepackage{algorithm}
\usepackage{array}
\usepackage{amssymb}
\usepackage[caption=false,font=normalsize,labelfont=sf,textfont=sf]{subfig}
\usepackage{textcomp}
\usepackage{stfloats}
\usepackage{url}
\usepackage{verbatim}
\usepackage{graphicx}
\usepackage{cite}
\usepackage{multirow}
\usepackage[figuresright]{rotating}
\usepackage{array}
\usepackage{xspace}
\usepackage{xcolor}
\usepackage{enumitem}
\usepackage{hyperref}

\newcommand{\eg}{\emph{e.g.,}\xspace}
\newcommand{\etal}{\emph{et al.}\xspace}
\newcommand{\etc}{\emph{etc.}\xspace}
\newcommand{\CO}{$\textrm{CO}_\textrm{2}$\xspace}

\begin{document}

\title{Advancing Sustainability via \\Recommender Systems: A Survey}
\author{Xin Zhou\IEEEauthorrefmark{1}, Lei Zhang\IEEEauthorrefmark{1}, Honglei Zhang, Yixin Zhang, Xiaoxiong Zhang, Jie Zhang, and Zhiqi Shen
\IEEEcompsocitemizethanks{
    \IEEEcompsocthanksitem Xin Zhou, Lei Zhang, Xiaoxiong Zhang, Jie Zhang and Zhiqi Shen are with the College of Computing and Data Science, Nanyang Technological University, Singapore. Honglei Zhang is with the School of Computer Science and Technology, Beijing Jiaotong University, China. Yixin Zhang is with the School of Software, Shandong University, China. (Email: xin.zhou@ntu.edu.sg, leizhzzl.1103@gmail.com, honglei.zhang@ bjtu.edu.cn, yixinzhang@mail.sdu.edu.cn, zhan0552@e.ntu.edu.sg, \{zhangj, zqshen\}@ntu.edu.sg.)
    \IEEEcompsocthanksitem $*$ Equal contribution.
    }
}

\maketitle
\begin{abstract}
Human behavioral patterns and consumption para- digms have emerged as pivotal determinants in environmental degradation and climate change, with quotidian decisions pertaining to transportation, energy utilization, and resource consumption collectively precipitating substantial ecological impacts. Recommender systems, which generate personalized suggestions based on user preferences and historical interaction data, exert considerable influence on individual behavioral trajectories. However, conventional recommender systems predominantly optimize for user engagement and economic metrics, inadvertently neglecting the environmental and societal ramifications of their recommendations, potentially catalyzing overconsumption and reinforcing unsustainable behavioral patterns.
Given their instrumental role in shaping user decisions, there exists an imperative need for sustainable recommender systems that incorporate sustainability principles to foster eco-conscious and socially responsible choices. This comprehensive survey addresses this critical research gap by presenting a systematic analysis of sustainable recommender systems. As these systems can simultaneously advance multiple sustainability objectives—including resource conservation, sustainable consumer behavior, and social impact enhancement—examining their implementations across distinct application domains provides a more rigorous analytical framework.
Through a methodological analysis of domain-specific implementations encompassing transportation, food, buildings, and auxiliary sectors, we can better elucidate how these systems holistically advance sustainability objectives while addressing sector-specific constraints and opportunities. Moreover, we delineate future research directions for evolving recommender systems beyond sustainability advocacy toward fostering environmental resilience and social consciousness in society.
\end{abstract}


\section{Introduction}
\begin{figure}[!t]
\centering
\includegraphics[width=0.48\textwidth, trim={10 10 10 0},clip]{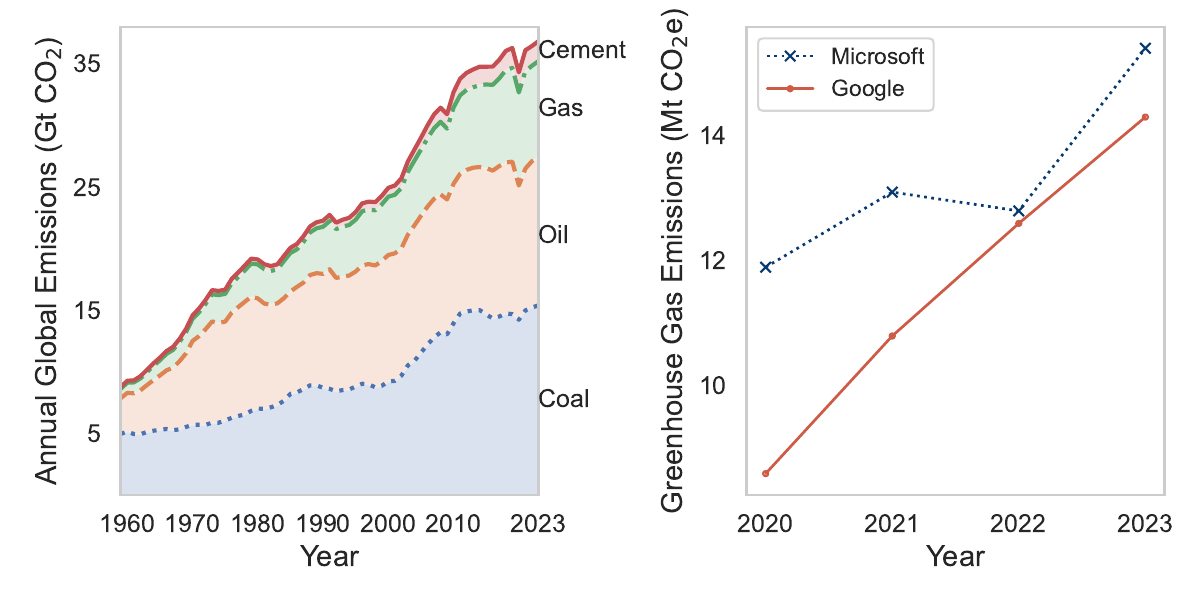}
\caption{Global \CO emissions trends (1960-2023) and a comparative analysis of Microsoft and Google's carbon footprints.}
\vspace{-12pt}
\label{fig_co2}
\end{figure}

{
\IEEEPARstart{H}uman activities encompass the consumption of myriad non-renewable resources (\eg coal, gas, fossil fuels) and natural materials, while concomitantly inflicting environmental degradation through various mechanisms, including atmospheric pollution, carbon dioxide (\CO) emissions, and waste generation~\cite{magnani2007human,salam2005impact,liu2020near}.
The cumulative global emissions of \CO have exhibited a consistent linear trajectory over time, as illustrated in Fig.~\ref{fig_co2} (left). According to the U.S. Energy Information Administration (EIA), the United States contributed approximately 5.12 billion metric tons (BMT) of \CO emissions in 2021~\cite{useia2023,wang2023simulating}. Of this total, a substantial 92 percent, or 4.6 BMT, was directly attributable to the combustion of fossil fuels for energy generation.
Moreover, recent years have witnessed a precipitous advancement in generative artificial intelligence (AI), which necessitates the utilization of large-scale datasets for training expansive language models~\cite{tomlinson2024carbon}.
A comparative analysis of the 2024 environmental sustainability reports from Microsoft and Google reveals a significant increase in their carbon footprints since 2020 (Fig.~\ref{fig_co2} (right)). Microsoft's emissions have risen by 29.4\%, while Google's have surged by 66.3\%. The primary driver of this growth is the expansion of their data center infrastructure, specifically designed and optimized to accommodate the escalating computational demands of artificial intelligence workloads.
It is incontrovertible that the magnitude of carbon emissions continues to escalate, making substantial contributions to anthropogenic climate change~\cite{milovanoff2020electrification,zhang2020role,kweku2018greenhouse}.
}

{
To align with the United Nations Sustainable Development Goals (SDGs)~\cite{fund2015sustainable} and adhere to the stipulations of the Paris Agreement aimed at mitigating climate change to below 1.5°C by the mid-2030s, a diverse array of artificial intelligence technologies has been deployed~\cite{kaack2022aligning}. As the global community becomes increasingly aware of the importance of sustainable practices to address environmental challenges, there is a growing consensus that recommender systems (RSs) can play a crucial role in facilitating sustainable human behaviors~\cite{FelfernigWTELMGL24,vente2024clicks}. These systems propose alternatives that endorse environmentally sustainable products, encourage eco-conscious travel options, and promote energy-efficient living arrangements within architectural structures.
Empirical research has demonstrated that recommender systems tailored for green products not only contribute to the diminution of energy usage and the reduction of greenhouse gas emissions~\cite{bosompem2024social} but also foster a paradigm of sustainable consumption among users. 
}

{RSs are AI models designed to predict user preferences and suggest relevant items or content. These systems have become ubiquitous in various domains, including e-commerce, entertainment, social media, and more~\cite{MMRec_Survey,CuiXXCCZC20,DRAGON23,GDCL22}. The primary goal of RSs is to enhance user experience by providing personalized recommendations tailored to individual preferences and behaviors~\cite{LiuSZXCYW23,LayerGCN}. Given the significant impact of RSs on daily human interactions with digital platforms, these systems have the potential to contribute substantially to environmental and social sustainability:
\textit{i)}. Waste reduction: By suggesting products or content that align more closely with user preferences, RSs can potentially mitigate waste from unwanted purchases or unused items.
\textit{ii)}. Energy efficiency optimization: In sectors such as energy management, RSs can propose optimal energy usage patterns, potentially promoting conservation and reducing carbon footprints.
\textit{iii)}. Promotion of sustainable consumption: These systems can be engineered to prioritize environmentally friendly products or services, potentially encouraging more sustainable consumer behavior.
\textit{iv)}. Enhancement of social well-being: Through the recommendation of educational content, health-related information, or community activities, these systems may contribute to social development and individual growth.
}

While several reviews have examined sustainable recommendation systems in specific domains, such as energy-efficient building practices~\cite{himeur2021survey}, eco-friendly travel routes~\cite{makhdomi2023towards}, sustainable e-tourism~\cite{rehman2021systematic}, and Sustainable Development Goals (SDGs) perspectives~\cite{FelfernigWTELMGL24}, this survey offers a more comprehensive and integrated analysis. The present study provides a holistic view of sustainable recommendation systems across multiple domains, including health-conscious food choices, energy-efficient building management, and environmentally friendly travel solutions, while also examining the underlying computational strategies employed in these systems. By synthesizing a diverse body of research and emphasizing the critical need to incorporate environmental sustainability into system designs, this review aims to enhance the understanding of sustainable recommender systems and stimulate future research that encompasses various aspects of sustainability. 
Our primary contributions are as follows:
\begin{itemize}[leftmargin=*]
\item We offer an in-depth examination of the implementation and research trajectories of sustainable recommendation systems in pivotal sectors, including travel, food, and built environment management, coupled with insights into algorithmic optimization. 
\item We present a generic architectural framework for sustainable recommender systems, serving as a foundation for organizing and contextualizing existing research.
\item We make a substantial contribution to the corpus of research on sustainable recommendation systems by elucidating key challenges within the domain and proposing future research avenues. It establishes a crucial framework for advancing the study and application of sustainability principles in recommendation systems across heterogeneous industries.
\end{itemize}

The subsequent sections of this paper are structured as follows: Section~\ref{sec:2} provides the necessary background for understanding the subsequent review. Section~\ref{sec:3} describes various work in sustainable travel recommendation, covering POI recommendation, route recommendation, and transportation recommendation. Section~\ref{sec:4} delves into sustainable recommendation practices within the food industry, focusing on health-conscious and environmentally friendly food recommendations. Section~\ref{sec:5} discusses the sustainable building recommendation, ranging from residential to commercial and large-scale buildings. Section~\ref{sec:6} expands to a wider discussion on the broader applications of sustainable recommendations, which includes environmental and ecological sustainability, behavior and social change, economic and productive sustainability, and user-centric sustainable recommenders. Section~\ref{sec:7} pivots to the sustainable design of recommendation models, particularly through algorithmic advancements and computational efficiency. Section~\ref{sec:8} highlights ongoing challenges and emerging research areas within the field. Finally, Section~\ref{sec:9} concludes the paper, synthesizing key findings and delineating implications for future research endeavors.

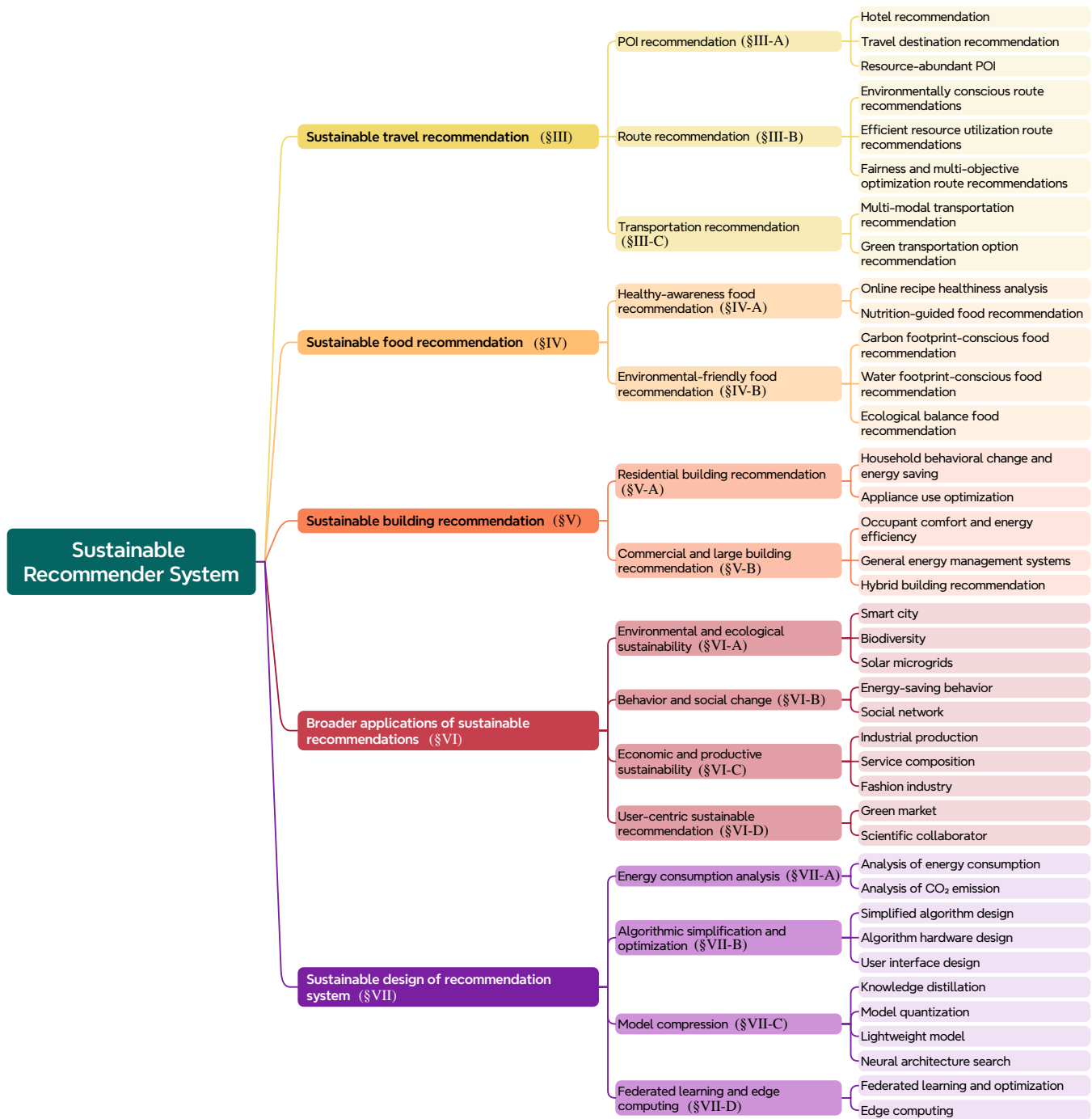
\begin{figure*}
\includegraphics[width=\textwidth, trim={1cm 4cm 1cm 1.3cm},clip]{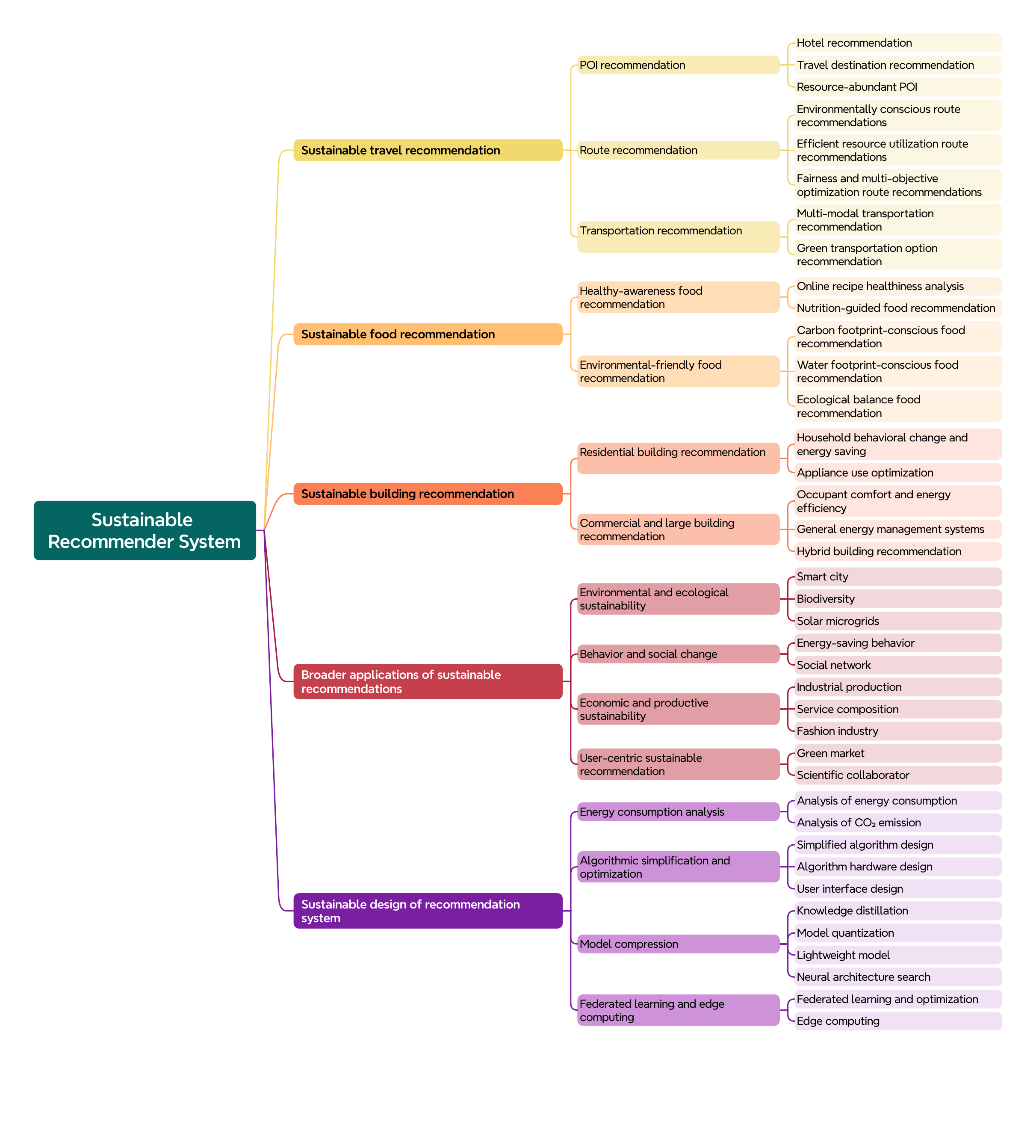}
\put(-262,467){\scriptsize (\S\ref{sec:3})}
\put(-168,512){\scriptsize (\S\ref{sec3:a})}
\put(-161,467){\scriptsize (\S\ref{sec3:b})}
\put(-224,418){\scriptsize (\S\ref{sec3:c})}
\put(-265,370.5){\scriptsize (\S\ref{sec:4})}
\put(-178,387){\scriptsize (\S\ref{sec4:a})}
\put(-178,348){\scriptsize (\S\ref{sec4:b})}
\put(-256,287.5){\scriptsize (\S\ref{sec:5})}
\put(-224,302){\scriptsize (\S\ref{sec5:a})}
\put(-178,265){\scriptsize (\S\ref{sec5:b})}
\put(-314,185){\textcolor{white}{\scriptsize (\S\ref{sec:6})}}
\put(-188,229){\scriptsize (\S\ref{sec6:a})}
\put(-151,203.6){\scriptsize (\S\ref{sec6:b})}
\put(-188,170.6){\scriptsize (\S\ref{sec6:c})}
\put(-178,142){\scriptsize (\S\ref{sec6:d})}
\put(-347,65){\textcolor{white}{\scriptsize (\S\ref{sec:7})}}
\put(-147,121.4){\scriptsize (\S\ref{sec7:a})}
\put(-190,88.8){\scriptsize (\S\ref{sec7:b})}
\put(-171,51.8){\scriptsize (\S\ref{sec7:c})}
\put(-193,13){\scriptsize (\S\ref{sec7:d})}
\caption{Hierarchical taxonomy of recommender system categories and their applications in advancing sustainability initiatives.}
\vspace{-12pt}
\label{fig:taxonomy}
\end{figure*}

\section{Background}
\label{sec:2}
\subsection{Sustainability}
{
The concept of sustainability has emerged as a paramount global imperative amid escalating challenges of climate change, resource depletion, and environmental deterioration. This paradigm was formally introduced through the seminal publication ``Our Common Future'', which established sustainability as the capacity to address present needs while preserving the ability of future generations to meet their requirements~\cite{hurlem1987our}. The conceptualization of sustainability, however, demonstrates significant contextual variability and exhibits stakeholder-dependent interpretations~\cite{vos2007defining}. This definitional flexibility underscores the complexity inherent in implementing sustainable practices across diverse domains.

In the contemporary artificial intelligence (AI) era, sustainability acquires additional dimensions of significance. While AI technologies present unprecedented opportunities for advancement and innovation, they simultaneously introduce substantial sustainability challenges that require careful consideration. Critical concerns include the exponential growth in energy consumption by data centers, the ethical ramifications of AI-driven decision-making processes, and the environmental footprint associated with large-scale AI infrastructure deployment.
As AI continues to evolve, it offers powerful tools to address sustainability challenges, such as optimizing resource utilization, reducing greenhouse gas emissions, and promoting sustainable practices across industrial sectors~\cite{vinuesa2020role}. 
Nevertheless, the accelerated progression of AI development necessitates rigorous evaluation of its long-term societal and environmental implications. The integration of sustainability principles into AI development and deployment frameworks represents a crucial imperative for safeguarding the future of our planet and the well-being of future generations.}

\subsection{Recommender Systems}
Recommender systems (RSs) are crucial technological solutions addressing the exponential growth of digital information accessibility~\cite{ricci2010introduction,MMRec_Lib,wu2022graph,DWSRec,bao2024improved,WhitenRec,MP4SR,hu2024mgdcf}. The proliferation of online content creates significant cognitive load for users navigating vast information spaces, necessitating efficient filtering mechanisms. These systems employ sophisticated algorithms to process and prioritize content, thereby optimizing information discovery and user engagement.

Recommender systems employ a variety of techniques to generate personalized suggestions for users, ranging from matrix factorization~\cite{koren2009matrix} to recent deep learning-based and graph-based recommender models~\cite{zhang2019deep,wu2022graph}.
{In the context of sustainability, RSs can serve as instrumental mechanisms for advancing sustainability objectives and promoting social welfare through the following critical pathways:
\begin{itemize}[leftmargin=*]
\item \textbf{Resource Conservation}: RSs minimize waste through preference-aligned product recommendations in high-consumption sectors including fashion and electronics, optimizing resource allocation and reducing disposal rates.
\item \textbf{Sustainable Consumer Behavior}: RSs facilitate behavior modification by prioritizing environmentally conscious and ethically produced goods within recommendation frameworks, enabling informed sustainable purchasing decisions.
\item \textbf{Social Impact Enhancement}: These systems optimize the dissemination of educational resources, social initiatives, and health-related information, contributing to collective societal well-being through targeted content distribution.
\item \textbf{Circular Economy Support}: RSs advance circular economy principles by facilitating secondary market transactions, sharing services, and recycling opportunities, effectively reducing primary resource demand through optimized resource allocation.
\end{itemize}
Given that recommender systems can concurrently advance multiple sustainability pathways, examining their implementations across distinct application domains provides a more comprehensive analytical framework. 
By analyzing RS implementations at the application level, including travel, food, buildings, and other domains, we can better understand how these systems holistically advance sustainability goals while addressing sector-specific challenges and opportunities. This methodological approach facilitates the investigation of multifaceted RS interventions that simultaneously address various sustainability pathways.
For instance, an intelligent transportation recommender system can optimize resource utilization through efficient routing while promoting environmentally conscious consumer behavior, thereby illustrating the intrinsic interconnectedness of sustainability pathways in practical implementations.
Hence, our review will expand at the application level of RS to advance sustainability.
Consequently, this review expands upon RS applications across different sectors to comprehensively examine their role in advancing sustainability. All surveyed publications are available in a curated GitHub repository at: \url{https://github.com/enoche/SusRec}.
}

\section{Sustainable Travel Recommendation}
\label{sec:3}
Travel recommendation systems aim to mitigate the daunting task of trip planning by providing personalized suggestions tailored to individual preferences~\cite{liu2011personalized,ge2011cost}. By harnessing user-specific behavior data, these systems offer insights and recommendations about users' journeys, ultimately enriching the overall travel experience. Specifically, travel recommendation is responsible for providing accurate points of interest (POI) recommendations (such as hotels, restaurants, \etc), efficient route recommendations, and appropriate transportation recommendations both before the commencement of the journey and during travel~\cite{cha2020trs,borras2014trs}. {However, traditional travel recommendation systems often neglect the environmental consequences of their suggestions, such as increased carbon emissions or ecological damage from promoting certain destinations, potentially leading to over-tourism and excessive resource consumption~\cite{BanerjeeMW24}.}

{As the environmental impact of tourism becomes more apparent, there is a growing need to integrate sustainability into these systems to encourage responsible travel choices.}
Sustainable travel recommendation (STR) is to furnish users with environmentally-friendly, resource-efficient, and time-saving travel services~\cite{makhdomi2023towards,Merinov24}. Specifically, it involves the incorporation of environmental preservation aspects concerning POIs~\cite{nilashi2019preference}, the promotion of low-carbon routes~\cite{garg2018route}, and sustainable transportation~\cite{makhdomi2023towards}. The ultimate goal of STR is to provide visitors with an all-encompassing green travel solution, involving destinations, routes, and modes of transportation. Consequently, this section will concentrate on three key aspects within the context of STR: POI recommendation, route recommendation, and transportation recommendation.

\subsection{POI Recommendation}
\label{sec3:a}
\begin{figure}
\centering
\includegraphics[width=0.48\textwidth]{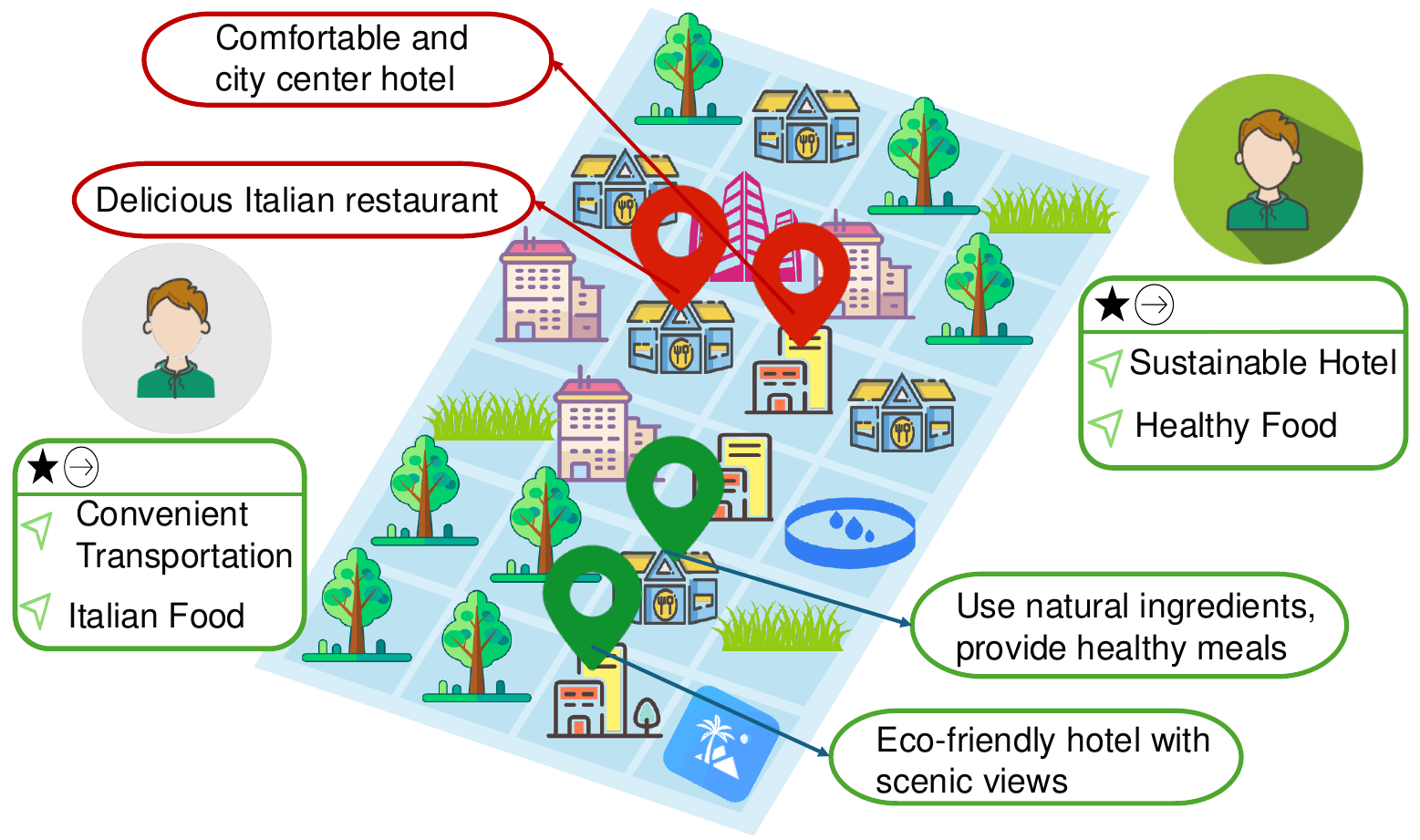}
\caption{An example of the sustainable POI recommendation.}
\vspace{-12pt}
\label{fig:poi}
\end{figure}

POI recommendation is the process of suggesting relevant future locations or points of interest to users based on their historical check-in data, preferences, and behaviors within location-based social networks, utilizing machine learning and data mining techniques to tailor these suggestions to individual user preferences~\cite{islam2022survey}.
It has been extensively studied by academia in recent years due to its advantages in many important aspects, including resource optimization and user experience personalization~\cite{sanchez2022point}. At the early stage, researchers first concentrated on traditional shallow models like Markov chain~\cite{zhang2014lore}, matrix factorization~\cite{lian2014geomf,cheng2013you}, and Bayesian personalized ranking~\cite{he2017category}. Recent research, however, has focused more on deep neural networks, such as LSTM~\cite{kong2018hst,GuiSYPLWGGG21}, RNN~\cite{yang2020location}, and self-attention models~\cite{XiaYWYCY24,he2017category}. Deep neural networks enhance the prediction of users’ future locations by effectively utilizing spatial-temporal correlations in mobility data like check-ins, significantly outperforming traditional POI recommendation models.

{Traditional POI recommendation methods, while effective at predicting users’ next points of interest, frequently overlook sustainability. There’s an increasing need to recommend POIs that not only are accurate but also align with green practices. This shift extends beyond merely suggesting popular or convenient locations to include those that support environmental conservation, local communities, and reduce ecological footprints. Consequently, sustainable POI recommendations encompass any suggestions that promote eco-friendly, socially responsible, and resource-efficient choices.}
For instance, it includes recommending environmentally-friendly hotels or eco-friendly public infrastructure~\cite{BanerjeeMW24,suanpang2022intelligent} as depicted in Fig.~\ref{fig:poi}.

As concerns for environmental sustainability rise and consumer awareness of environmental issues grows, eco-friendly services such as sustainable POI recommendations have attracted significant attention~\cite{nilashi2019preference,suanpang2022intelligent}. For example, Nilashi \etal~\cite{nilashi2019preference} proposed a multi-criteria collaborative filtering model for eco-friendly hotel recommendations, utilizing machine learning and dimensionality reduction techniques to enhance scalability in predicting traveler preferences based on a large TripAdvisor dataset, demonstrating robust performance in forecasting user choices. Suanpang \etal~\cite{suanpang2022intelligent} introduced a spatio-temporal multi-agent reinforcement learning framework to intelligently recommend public-accessible charging stations, accounting for various long-term factors. Experimental results indicate the framework’s effectiveness in reducing average charging costs, failure rates, and wait times for electric vehicles, thereby promoting low-carbon, eco-friendly travel through sustainable POI recommendations. {Besides, Banerjee \etal~\cite{BanerjeeMW24} proposed the Green Destination Recommender, a web application leveraging React to recommend sustainable travel destinations by integrating \CO emissions, popularity, and seasonality indices, aiming to encourage environmentally conscious travel choices.}

The user-centric approach of traditional tourism industry POI recommendation methods can lead to long-term negative impacts on the environment and local communities. This strategy fosters user-driven tourism, which adversely affects wildlife habitats and various landscapes where tourism takes place~\cite{bramwell1993sustainable}. Furthermore, this approach could also detrimentally affect the experiences of both tourists and locals. For instance, crowded tourist areas might result in reduced social distances, increased noise levels, and traffic congestion. In order to lessen the detrimental effects of unchecked development in the name of economic gain, Merinov~\cite{merinov2023sustainability} intentionally attempted to ease the tensions brought on by the intricate relationships between visitors and the environment.

Besides recommending specific sustainable POIs (such as hotels or charging stations), another research direction is exploring capacity constraints in POI recommendation~\cite{christakopoulou2017recommendation,merinov2023sustainability}. In other words, providing users with resource-abundant POIs can indirectly protect the environment, while recommending resource-limited POIs can easily lead to overcrowding and unnecessary resource waste. A maximum capacity, such as the number of seats in a POI or the quantity of item copies in the inventory, is often linked to the candidate items for POI recommendations. To address this issue, Konstantina \etal~\cite{christakopoulou2017recommendation} proposed a multi-objective framework that simultaneously optimizes recommendation accuracy and the capacity constraints of the recommended items. By providing users with real-time information on POI capacity availability, the framework effectively guides users to appropriate locations, achieving the goal of offering sustainable POIs from an algorithmic perspective. The effectiveness of this method has been validated on three classic matrix factorization methods: PMF~\cite{mnih2007probabilistic}, BPR~\cite{rendle2012bpr}, and GeoMF~\cite{lian2014geomf}. Besides, Merinov~\cite{merinov2023sustainability} addressed POI recommendation by mitigating popularity bias, ensuring a healthy distribution of visitor traffic. This approach promotes the long-term economic viability of the environment.

\subsection{Route Recommendation}
\label{sec3:b}
Route recommendation plays a crucial role in personalized travel services by providing users with reasonable and scientifically travel routes based on their historical behavior data and user profiles~\cite{dai2015personalized}. For drivers, these systems identify routes with a higher likelihood of finding passengers, reducing idle cruising time and increasing profits. For passengers, these recommendations aim to improve satisfaction by directing drivers along the most efficient paths, minimizing waiting times, and ensuring a quicker arrival at their destinations.

Recent academic research on route recommendation has focused on two primary approaches: conventional route planning and personalized route recommendation~\cite{yuen2019beyond,garg2018route}. Conventional route planning involves classic shortest path algorithms like Dijkstra’s and A*, with advancements that account for variables such as uncertain travel times~\cite{hua2010probabilistic}. Personalized route recommendations, on the other hand, include methods like TRIP~\cite{letchner2006trip}, which relies solely on individual driving data, T-drive~\cite{yuan2010t}, which uses driving time for recommendations, and more advanced methods~\cite{dai2015personalized} that integrate comprehensive driving information for enhanced efficiency.

While route recommendation systems have greatly enhanced people's lives and productivity, mounting evidence of their detrimental effects on the environment has compelled some cities to address the urban mobility issues. Route planning is often blamed for contributing significantly to pollution by creating traffic congestion and leading to situations where drivers roam without passengers. It is crucial to develop sustainable route recommendation systems that reduce fossil fuel consumption, thereby minimizing local air pollution and environmental damage~\cite{sengvong2017persuasive}, as illustrated in Fig.~\ref{fig:route} (a). For instance, Bothos \etal~\cite{bothos2012recommending} introduced an eco-friendly methodology for travel recommendation systems, proposing a system structure that merges profile matching techniques with essential information elements. This system offers users alternative routes that align with their preferences while also reducing carbon emissions. Additionally, Makhdomi \etal~\cite{makhdomi2023towards} explored fairness and environmental issues on ride-hailing platforms and provided a comprehensive overview of advancements in route recommendation for these services. Furthermore, Bao \etal~\cite{bao2017planning} leveraged bike-sharing trajectory data to design efficient bike lanes, enhancing safety and encouraging eco-friendly travel.
Sotsay \etal~\cite{sengvong2017persuasive} introduced a route recommendation system that helps commuters make safer, eco-friendly, and less congested travel choices while reducing societal costs like accidents and pollution. This system employs a persuasive reward algorithm and an agent-based model to evaluate the effectiveness of recommendations, achieving a higher public-friendly score than traditional methods. Additionally, Namoun \etal~\cite{namoun2021eco} introduced an eco-friendly multi-modal route recommendation system that simulates complex urban transportation networks with software agents. It leverages real-time traffic data, including carbon emissions and flow patterns from diverse sources, to offer travelers optimized route suggestions via a dedicated application.

The aforementioned methods achieve sustainable route recommendation by directly considering environmental factors. For example, they incorporate fairness into route recommendation or consider \CO emissions and congestion levels during modeling, significantly enhancing the sustainability of route recommendations. Besides explicitly modeling environmental features, some studies explore efficient shortest path recommendation methods to reduce carbon emissions, indirectly contributing to the vision of sustainable route recommendation~\cite{yuen2019beyond,garg2018route}. 
Specifically, Garg \etal~\cite{garg2018route} focused on reducing the distance between idle taxis and upcoming customers to enhance driver productivity and minimize customer wait times. They created a route recommendation engine, MDM (Minimizing Distance through Monte Carlo Tree Search), which forecasts the likely origins of future customer requests to suggest efficient routes. Besides, Yuen \etal~\cite{yuen2019beyond} investigated taxi ride-sharing by developing the Share algorithm, which optimizes routes to accommodate multiple passengers without significant detours. This method effectively reduces the need for multiple taxis, lowering carbon emissions and enhancing sustainable transport. 

Furthermore, designing memory-efficient and time-saving mechanisms for route recommendation models also contribute to sustainability goals. Qu \etal~\cite{qu2014cost} employed a Brute-Force method to derive optimal driving routes, crafting a graph representation of road networks from historical taxi GPS data. To counter the high computational cost of this graph-based approach, they introduced a recursive technique that efficiently identifies the most profitable routes. The aforementioned methods enhance travel efficiency and user satisfaction but also prompt environmental concerns, driving the need for sustainable solutions that reduce pollution and promote eco-friendly travel practices.

\subsection{Transportation Recommendation}
\label{sec3:c}
\begin{figure}[tb!]
\centering
\begin{minipage}[t]{0.5\linewidth}
\centering
\includegraphics[width=\textwidth]{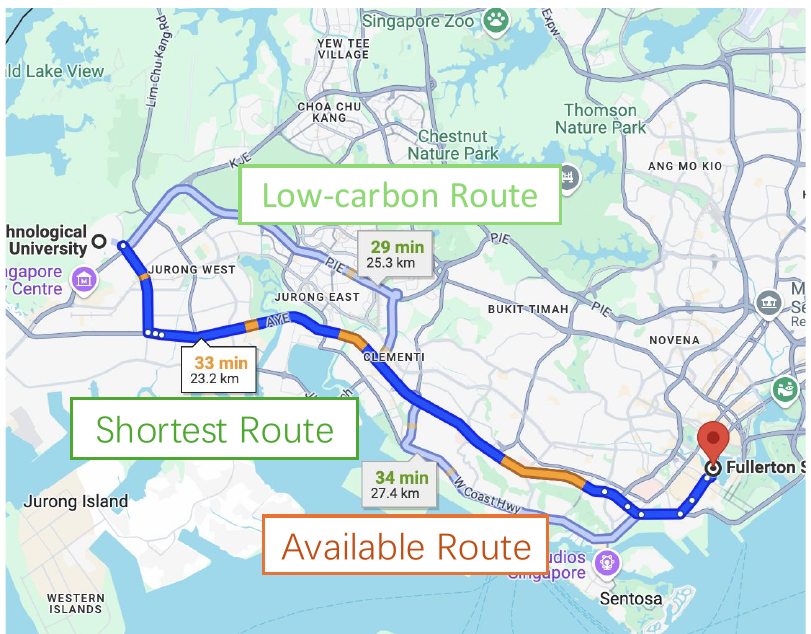}
\centerline{(a) Route}
\end{minipage}%
\begin{minipage}[t]{0.5\linewidth}
\centering
\includegraphics[width=\textwidth]{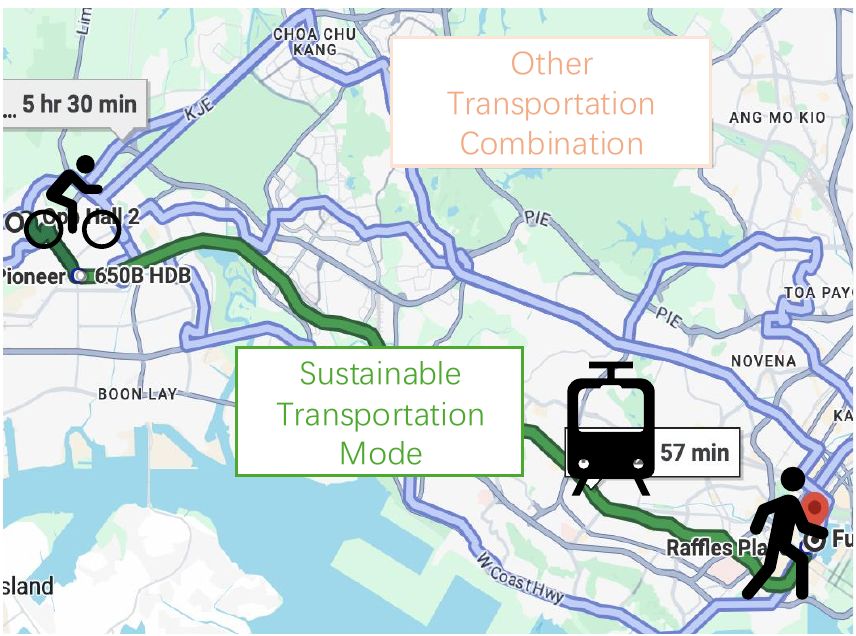}
\centerline{(b) Transportation}
\end{minipage}%
\centering
\caption{Illustration of sustainable route and transportation recommendations.}
\vspace{-12pt}
\label{fig:route}
\end{figure}

Transportation modes are ways to move from an origin to a destination, including driving, walking, cycling, and public transportation~\cite{liu2019joint}. The proliferation of different modes of transportation (bus, car, ride-sharing, shared-bike, \etc) and the swift growth of transportation networks (pedestrian, bus, road, \etc) have given travelers an abundance of options to reach their destinations~\cite{liu2020multi}. Transportation recommendation refers to the task of identifying the best transportation options while taking into account trip attributes (such as destination and distance) and user preferences (such as habits and times), as exemplified in Fig.~\ref{fig:route} (b). Transportation recommendations provide reasonable and timely travel options for tourists, offering a convenient experience, and are widely applied in many internet applications, such as Google Maps, Baidu Maps. 

Classic transportation recommendation methods are typically divided into two categories: single-mode planning and multi-modal recommendation~\cite{chaudhari2020comprehensive}. Early research focused on single-mode transport, such as using trajectory data for private vehicle planning~\cite{dai2015personalized}, and leveraging mobility data to suggest popular routes between locations~\cite{chen2011discovering}. However, real-world planning often requires choosing the best combination of multiple transport modes, leading to increased attention on multi-modal recommendations~\cite{liu2020multi}. Multi-modal approaches are categorized into retrieval-based methods, which use predefined metrics and graph algorithms like Dijkstra~\cite{borole2013multimodal,dibbelt2016engineering}, and learning-based methods. They apply machine learning to derive transport preferences or improve stability using features like estimated time of arrival~\cite{liu2019hydra,liu2019joint,IGT23,zhang2024estimating,zhang2023dual,zhang2023delivery}.

Although the aforementioned methods effectively combine various modes of transportation, allowing tourists to travel smoothly from their point of origin to their destination, they often solely focus on reducing travelers' waiting and travel times, neglecting the multiple goals of sustainable development and environmental protection~\cite{dubey2015sustainable}. Consequently, the emergence of sustainable transportation recommendations has been crucial in addressing these oversights. These recommendation systems aim to ensure travelers reach their destinations smoothly and on time while also prioritizing environmental protection~\cite{makhdomi2023towards}. Technically, sustainable transportation recommendations need to consider the reasonable combination of multiple transportation modes and the shortest path search with carbon emission constraints~\cite{bothos2012recommending}. For example, Arnaoutaki \etal\cite{arnaoutaki2021recommender} presented a hybrid knowledge-based recommender system designed to support Mobility-as-a-Service (MaaS) users in selecting mobility plans tailored to their specific transportation needs and preferences. Additionally, recommending green transportation options to travelers~\cite{ge2010energy,yuan2023policy} is the effective approach to sustainable transportation recommendations.

Regarding green transportation alternatives, many cities have recently adopted bike-sharing systems to foster eco-friendly travel and promote healthier living~\cite{cheng2018utilization}. The placement of bike stations strategically to best suit users' trip demands is a critical component in optimizing the effectiveness of these systems. Urban planners typically use in-depth surveys to determine the demand for local bike trips. But this method is labor- and time-intensive, especially when comparing multiple possible locations. To address these issues, Chen \etal~\cite{chen2015bike} formulated the problem of bike station placement as a demand prediction problem for bike trips. In order to predict the demand for bike trips, they propose using a semi-supervised feature selection method to extract customized features from highly variable, heterogeneous urban open data. 

In addition to directly recommending green transportation options, Bothos \etal~\cite{bothos2012recommending} developed a travel recommendation system aimed at encouraging eco-friendly habits among environmentally conscious travelers. The system offers personalized advice on the greenest transportation options by incorporating profile matching techniques and recommendation information into its architecture. Besides, Ge \etal~\cite{ge2010energy} conducted a study on how to harness location data to foster energy-efficient transportation. They developed a mobile recommendation system that suggests optimal parking spots or pickup locations for taxi drivers based on energy-saving travel patterns identified from location traces. This system not only optimizes operational efficiency but also promotes sustainable driving behaviors among drivers, enhancing their chances of business success.

Xu \etal~\cite{xu2022machine} proposed a framework to enhance recommendation systems for connected and autonomous vehicles using hidden relationship mining and dual joint matrix factorization models. These models, driven by machine learning, efficiently process overlooked data to refine vehicle connectivity recommendations. In summary, integrating green transportation options like bike-sharing and carpooling into transportation recommendation systems, while emphasizing low-carbon and environmentally friendly criteria, represents a strategic approach to fostering sustainable travel. Such practices not only cater to the evolving preferences of travelers but also align with global sustainability goals, ultimately contributing to a reduction in environmental impacts and promoting healthier living standards.

\section{Sustainable Food Recommendation}
\label{sec:4}

\begin{figure}
\centering
\includegraphics[width=0.48\textwidth]{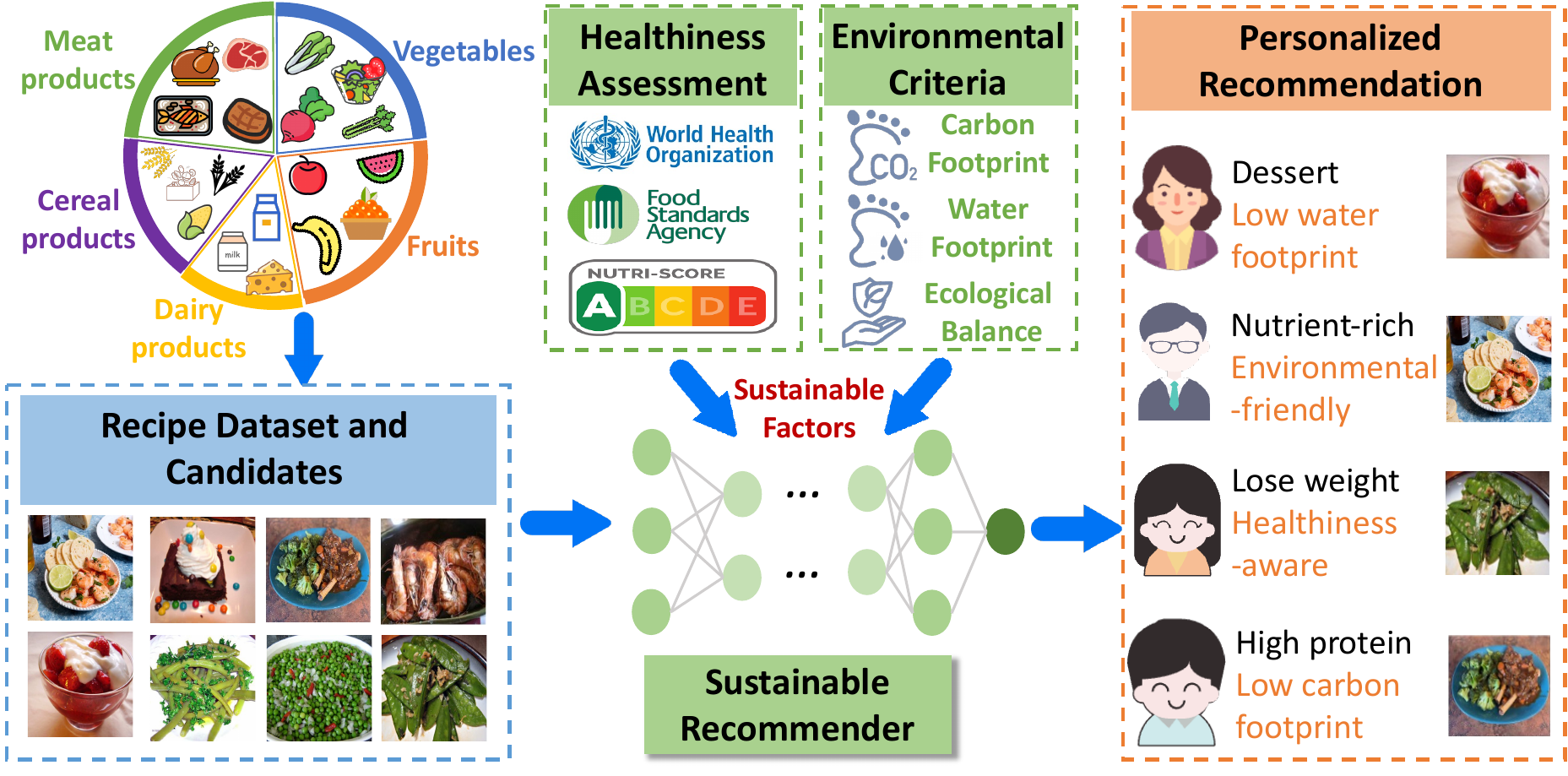}
\caption{Illustration of the sustainable food recommendation.}
\vspace{-12pt}
\label{fig:food}
\end{figure}

Food recommendation systems~(FRSs)~\cite{food_survey} serve as pivotal components in the realm of digital lifestyle services, designed to assist users in discovering recipes and food items that resonate with their unique dietary predilections. In today's interconnected world, where information is abundant and diverse, FRSs act as intermediaries that guide users towards food choices that are not only aligned with their taste preferences but also conducive to their health and well-being. These systems utilize a variety of data sources, including user preferences, dietary restrictions, and nutritional needs, to provide tailored recommendations, as shown in Fig. \ref{fig:food}.

Traditional FRSs primarily focus on leveraging user interaction data, multi-modal content, ingredient information, \etc, to personalize recipe suggestions that align with individual dietary needs. 
For example, Ge \etal~\cite{food_mf} proposed an extension method using tags within a matrix factorization framework to enhance prediction accuracy. Furthermore, Trattner \etal~\cite{LDA} demonstrated that Latent Dirichlet Allocation (LDA) and weighted matrix factorization methods outperform other collaborative filtering techniques, with research showing that incorporating item-side information like recipe ingredients and images effectively addresses data sparsity and precisely models user preferences. In this vein, Gao \etal~\cite{HAFR} introduced a hierarchical attention mechanism that concurrently integrates user-recipe interactions, recipe images, and ingredients for improved recommendation.
Moving beyond general collaborative filtering methods, recent research has explored leveraging Graph Neural Networks (GNNs) for food recommendation. Gao \etal~\cite{FGCN} utilized a graph convolution network to capture the intricate relationships between ingredients, recipes, and users. Similarly, Tian \etal~\cite{HGAT} constructed heterogeneous recipe graphs to model the connections between recipe content and its structure. Tian \etal~\cite{RecipeRec} introduced a graph contrastive augmentation strategy based on the user-recipe-ingredient heterogeneous graph to extract informative graph knowledge in a self-supervised manner. Meng \etal~\cite{PiNet} tackled both ingredient prediction and food recommendation tasks simultaneously, allowing it to learn visual features that incorporate both semantic and collaborative information. Zhang \etal~\cite{CLUSSL} employed clustering and self-supervised learning to exploit multi-modal recipe information for recommendation.

\subsection{Healthy-awareness Food Recommendation}
\label{sec4:a}
As the Internet and information technology continue to advance, there has been a marked improvement in the cognitive capabilities of the populace. This has led to a growing recognition of the critical role that dietary intake plays in shaping human health and well-being, as well as promoting sustainable living through its impact on both physiological and psychological conditions. Specifically, Food recommendation systems are becoming indispensable tools in promoting healthier eating habits, enhancing the overall quality of life, and assisting users in achieving specific health goals, including weight management, disease prevention, and maintaining nutritional balance.

Although there is currently no unified standard for measuring the healthiness of recipes or foods, numerous researchers have conducted exploratory studies based on the suggestions for daily nutrient intake provided by the World Health Organization (WHO)~\cite{WHO} and the UK's Food Standards Agency (FSA)~\cite{FSA}, as well as a five-coloured label system Nutri-Score~\cite{nutri-score} developed by Nutritional Epidemiology Research Team. These studies primarily assess the healthiness of recipes by analyzing the alignment of the nutritional components contained in them, such as calories, fats, sugars, salt, vitamins, \etc, with these health standards. 

According to the nutritional guidelines published by WHO and FSA, Trattner \etal~\cite{health_casestudy} analyzed the healthiness of main meal recipes from online websites, celebrity chef cookbooks, and ready meals from supermarkets. Their analysis suggests that online recipes are less likely to meet these standards and tend to be less healthy. Further research by Trattner \etal~\cite{LDA} revealed that the recipes on these websites are quite unhealthy, and users generally interact most frequently with the least healthy recipes. Although most recipe websites prioritize popular recipes, which tend to be unhealthy, Starke \etal~\cite{choices-visual} utilized visual enhancements of images and re-ranking of results on recipe websites to support healthier food choices without decreasing user satisfaction.

WHO defines 15 nutrient ranges crucial for a balanced daily meal plan. The existing research~\cite{HFRS, LDA} mainly focuses on the 7 most significant nutrients: proteins, carbohydrates, sugars, sodium, fats, saturated fats, and fibers. Each food is assigned a healthiness score according to the criteria, and the scale ranges from 0 to 7 (0 meaning none of the WHO ranges are fulfilled and 7 meaning all ranges are met). Consequently, a food item scoring 7 (meeting all WHO-defined ranges) is considered exceptionally healthy, while a score of 0 (failing to meet any WHO-defined ranges) is considered highly unhealthy. Research has shifted towards incorporating health criteria into recommendation algorithms. Forouzandeh \etal~\cite{HFRS} utilized node-level and semantic-level attention mechanisms to identify recipes that are both popular and healthy, then recommend healthy meals based on a heterogeneous graph and the dual attention mechanism. Rostami \etal~\cite{health_time} developed an efficient food recommendation method based on user preference and healthy nutrition factors, and introduced a novel time-aware similarity function to incorporate the dynamic nature of users’ preferences. Ahmadian \etal~\cite{health_reliablity} introduced an effective health-aware reliability measurement, which simultaneously considers both the accuracy and the health factors of predicted ratings to evaluate their reliability value. The proposed method incorporates health-aware reliability measurement into time-aware food recommendation.

FSA proposes a green, orange, and red traffic light system to evaluate the healthiness of recipes and provides standard ranges for low (green), medium (orange), and high (red) content of fat, saturates, sugar, and sodium. To calculate a health score, one point is awarded if an element’s quantity is within the low range, two for the medium range, and three for the high range, resulting in a health score ranging from 4 (best) to 12 (worst). Pecune \etal~\cite{fsa_1} indicated that in a hybrid recommendation system that combines personalization with health awareness, users who pay attention to the health tag are more likely to choose the healthy recipes recommended by the system, emphasizing the importance of accurately inferring individuals’ dietary goals and customizing the recommendation algorithms accordingly. Another observation is that people tend to avoid recipes labeled as unhealthy (red) as well as those labeled as healthy (green); the former due to feelings of guilt, while the latter is perceived as ``healthy equals less tasty''. Rostami \etal~\cite{health_fairness} explored health-aware and fairness-aware food recommendation, proposing a re-ranking method that integrates user and recipe fairness constraints.

Nutri-score is a 5-colour nutrition label (A, B, C, D, E) rating the overall nutritional value of food products. Among them, A represents the healthiest option, while E represents the least healthy. Nutri-score aims to provide nutritional information about food through a simple color-coded label, helping consumers quickly understand the nutritional content of food and make healthier choices. This system was initially developed in France and has been promoted and used in several European countries. Nurbakova \etal~\cite{nutri-colour} explored incorporating the recipe Nutri-score as an additional constraint into the KBQA framework based on the knowledge graph HUMMUS~\cite{HUMMUS} to develop a system capable of providing adaptive and personalized health food recommendations. 

In addition, numerous aspects of nutrition can be considered when recommending recipes. Ng \etal~\cite{nutri_1} proposed a personalized recipe recommendation system for toddlers that combines standard nutrition guidelines from the U.S. government’s ChooseMyPlate with users’ food preferences. Bianchini \etal~\cite{medical_food_health} provided users with personalized and healthy food options by taking into account both their preferences and medical prescriptions. Pallavi \etal~\cite{caloier} employed the Harris-Benedict equation to estimate users’ basal metabolic rate (BMR) and integrate their daily calorie intake requirements into the recommendation system. Song \etal~\cite{SCHGN} considered users’ evolving preferences for caloric intake by constructing a heterogeneous graph that models the complex relationships between users, recipes, ingredients, and calorie information. Zioutos \etal~\cite{share} combined a content-based and collaborative filtering approach, introducing a knowledge-based component to analyze nutritional information from recipes, thereby tailoring recommendations to suit users’ chronic health conditions. Zhang \etal~\cite{HealthRec2024} distills knowledge from the metadata of recipes, including descriptions, ingredients, and health-related attributes, to enhance food recommendation systems.
Moreover, a survey involving 40 real users showed that integrating health information with personalized filtering more closely aligns with user requirements. Finally, Zhang \etal~\cite{GreenRec} facilitated further research by releasing the first-ever green food dataset with integrated health scores. 

\subsection{Environmental-friendly Food Recommendation}
\label{sec4:b}
With the rise of digital literacy and accessibility to information, there is a growing awareness among users about the importance of dietary intake and its influence on both lifestyle and society. This awareness is not limited to personal health alone, there is a burgeoning recognition of the environmental impact of food choices. Modern consumers are increasingly considering sustainability factors such as carbon footprint, water footprint (WF), and ecological balance when making food choices. Consequently, there is a rising demand for food recommendation systems that can accommodate these complex preferences and provide suggestions that are not only personalized but also environmentally responsible.

In food recommendation systems, carbon footprints serve as a critical metric for evaluating and optimizing dietary choices to reduce environmental impact while maintaining nutritional balance. Gonz{\'a}lez \etal~\cite{gonzalez2018carbon} found that a diet rich in vegetables has a lower carbon footprint than one rich in meat. While reducing the intake of animal products may benefit the environment, it could also limit the intake of certain nutrients. It is crucial to develop consistent and widely accepted methods for estimating carbon footprints and assessing nutrient quality scores in diet recommendations. 

Water footprint is a burning contemporary problem in terms of sustainability. Mekonnen \etal~\cite{green_blue_grey} quantified the green, blue, and grey WF of global crop production from 1996–2005. Green WF measures rainwater consumption in production; blue WF quantifies surface and groundwater used; grey WF assesses freshwater needed to dilute pollutants to acceptable levels. A recipe’s WF includes both direct and indirect water required for its production. Blas \etal~\cite{FR_water_evaluate} used the Water Footprint Assessment (WFA) to measure the green, blue, and grey water footprints of diets, finding that a Mediterranean diet significantly conserves water, emphasizing the need for improved production and consumption to enhance environmental sustainability. Sobhani \etal~\cite{towards_WF} assessed the usual food intake of 723 individuals, aged 20 to 64 years, from Urmia, Iran. Based on this survey, linear programming techniques are used to identify an optimal dietary pattern that considers WF reduction. This study underscores the significance of sustainable diets, highlighting the potential for water conservation through modifications in dietary practices. Tompa \etal~\cite{tompa2020sustainable} also emphasized the importance of adopting sustainable eating patterns that not only enhance people’s health but also effectively reduce the environmental burden, particularly by decreasing the WF. Gallo \etal~\cite{FR_water} employed ingredient categorization and corresponding water footprint (WF) values to estimate the WF of entire recipes by summing the WF of each ingredient. This approach enables an assessment of the impact of various recipes on water resources. Additionally, a recommender system that analyzes and predicts users' dietary habits can provide personalized food suggestions aimed at reducing water footprints, promoting more environmentally sustainable and health-conscious meal choices.

Ecological balance functions as a guiding principle to ensure that suggested diets not only meet nutritional needs but also support the sustainable use of natural resources and biodiversity conservation. Irz \etal~\cite{irz2016welfare} provided a detailed description of the effects of different dietary recommendations on food consumption, nutrition, and environmental indicators, along with cost-benefit analyses. They found that a combination of health and environmental concerns is necessary to ensure consistency in consumers’ dietary advice, noting that reducing meat consumption and shifting to a plant-based diet are beneficial for both the environment and health. The study calls for continued efforts to promote dietary guidelines that support sustainability. Behrens \etal~\cite{behrens2017evaluating} argued that food systems impose significant environmental burdens, including global greenhouse gas emissions, eutrophication, and land use associated with food production. According to the survey, nationally recommended diets (NRDs) in high-income countries generally recommend reducing the intake of sugar, oil, meat, and dairy products, replacing these with fruits, vegetables, and nuts to provide additional nutrients. In contrast, NRDs in low-income countries are more likely to focus on increasing caloric and protein intake, particularly to address malnutrition and micronutrient deficiencies, by encouraging the consumption of more meat and fish. The production of animal-based products often carries higher environmental costs, including increased greenhouse gas emissions, greater land use, and heightened eutrophication. Based on these findings, they suggest that environmental sustainability needs to be considered in optimizing NRDs. Perignon \etal~\cite{perignon2019meet} proposed estimating the environmental impact of food items using seven metrics: water deprivation, land use, land use potential impacts on erosion resistance, mechanical filtration, groundwater replenishment, biotic production, and biodiversity. Furthermore, they reveal the necessity to strike a balance between nutritional and environmental objectives when formulating dietary recommendations and propose methods to achieve this by optimizing food choices in the Mediterranean region, particularly in Tunisia.

\section{Sustainable Building Recommendation}
\label{sec:5}
\begin{figure}
\centering
\includegraphics[width=0.48\textwidth]{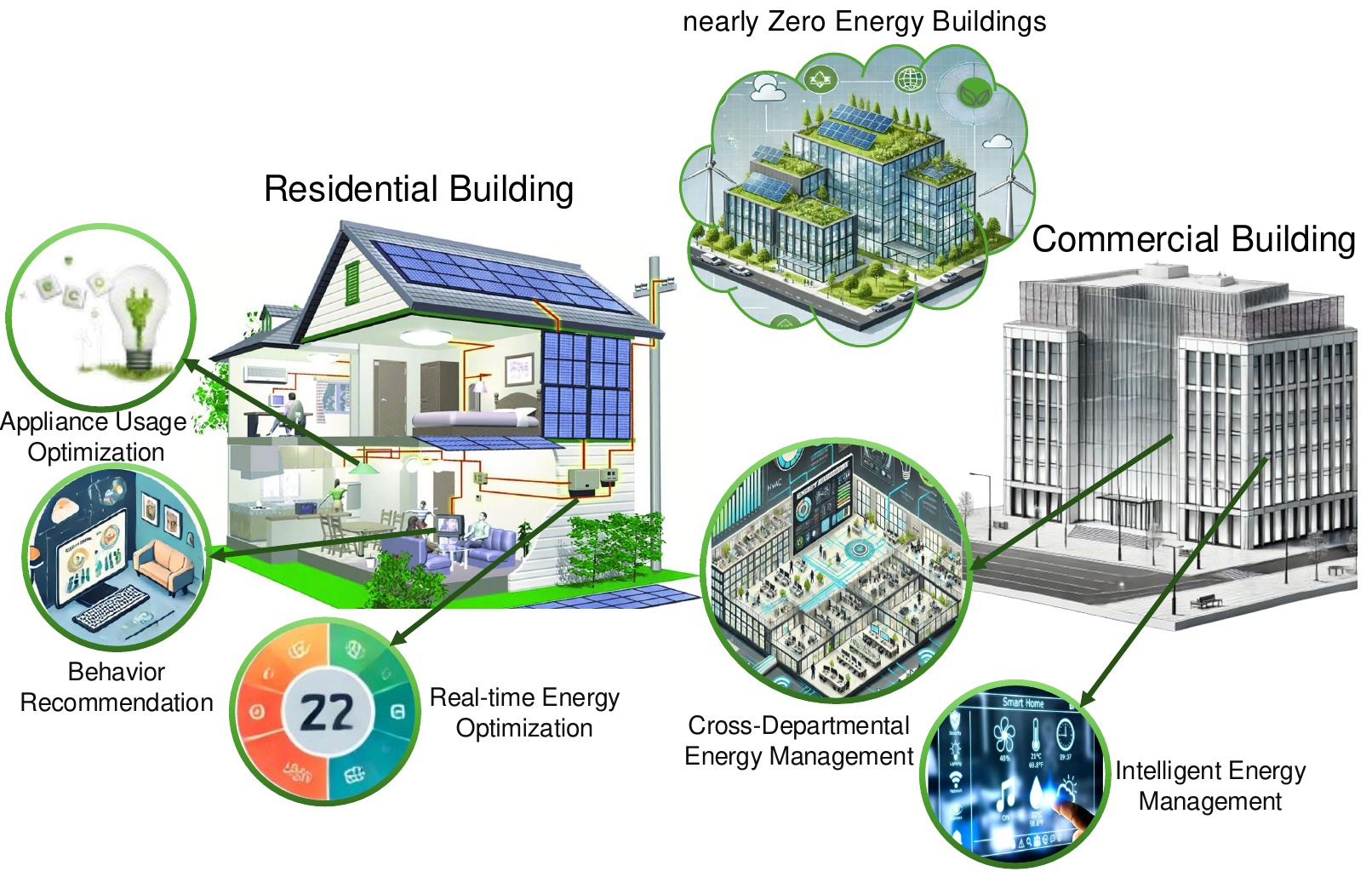}
\caption{Example of sustainable building recommendation.}
\vspace{-12pt}
\label{fig:building}
\end{figure}

Building recommendation systems are increasingly pivotal in the construction and facility management sectors, crucial for optimizing building performance, enhancing comfort, and cutting operational costs. As buildings grow more complex and demand more energy, the importance of deploying intelligent, sustainable systems to efficiently manage and optimize these structures becomes more pronounced. Energy consumption is a critical issue in the building sector, which is responsible for over 40\% of global energy use~\cite{cao2016building}. Projections indicate that this consumption will grow significantly, with an average annual increase of around 1.3\% from 2018 to 2050, and rates exceeding 2\% in some regions~\cite{economidou2020review}. Current research in sustainable building recommendations is primarily focused on two key areas: residential buildings and commercial and large buildings, as shown in Fig.~\ref{fig:building}. The key distinction between residential and commercial building recommendations hinges on their focus areas: residential strategies emphasize individual household energy optimization and behavior adjustments, while commercial strategies tackle extensive energy management and occupant engagement across larger infrastructures.

\subsection{Recommender Systems in Residential Buildings}
\label{sec5:a}
Research on residential building applications focuses on developing recommendation systems that enhance energy efficiency by targeting household energy consumption patterns and behaviors~\cite{DEirinakiVDJPT22}. In residential settings, energy use is largely influenced by occupants’ daily routines and habits, leading to significant variability in consumption. Hence, it is vital to provide sustainable recommendations that help residents seamlessly integrate energy-saving practices into their daily lives, enhancing their ability to effectively manage appliance use. This approach not only reduces energy waste but also ensures that comfort and convenience are maintained, addressing the growing need for sustainable living solutions in response to increasing energy demands and environmental concerns.

A significant portion of the literature on household energy savings focuses on developing recommendation systems that encourage behavioral changes to reduce energy consumption. These systems leverage data analytics and machine learning techniques to analyze user behaviors and provide tailored recommendations that seamlessly integrate energy-saving actions into daily routines. For example, Sardianos \etal~\cite{sardianos2019want} proposed a context-aware recommendation system that uses association rule mining to identify energy consumption patterns and provide personalized suggestions, such as adjusting heating settings based on contextual data like temperature, to promote gradual behavior changes. Dahihande \etal~\cite{dahihande2020reducing} developed a real-time recommendation system designed to optimize household energy usage. The system utilizes historical consumption data to identify abnormal appliance use, offering real-time and tailored advice that leads to substantial energy savings. Similarly, Pruvost \etal~\cite{pruvost2022recommendation} developed a rule-based system that continuously monitors real-time building data to deliver customized energy-saving strategies, helping users make informed decisions to reduce energy consumption effectively. Besides, Starke \etal~\cite{UMAPStarkeWS21} introduced the ``Saving Aid'' recommendation system for household settings, aiming at promoting energy-saving behaviors through two innovative framing mechanisms: the ``Savings Score'', which ranks recommendations based on kWh savings, and the ``Smart Savings Score'', which combines these savings with the perceived effort required. This approach underscores the importance of framing and presentation in encouraging sustainable behaviors in households through personalized recommendations.

Some studies focus on optimizing appliance usage to minimize energy consumption by using advanced algorithms that consider electricity prices, user preferences, and historical usage patterns to identify the most energy-efficient operation times. Jimenez \etal~\cite{jimenez2019multi} developed a multi-agent recommendation system that analyzes usage data from household devices and current electricity prices to offer personalized usage recommendations. This system helps users optimize appliance use times, thereby reducing peak electricity consumption and saving costs. Luo \etal~\cite{luo2020personalized} developed a collaborative filtering-based system that offers personalized energy-efficient appliance-use recommendations by drawing on the experiences of individuals with similar lifestyles while smarter appliance habits. Besides, Riabchuk \etal~\cite{riabchuk2024utility} developed a utility-based, context-aware system for load management in residential buildings. It models user preferences based on availability and device usage patterns, incorporating electricity price forecasts and historical consumption data to suggest optimal appliance start times for the following day, ensuring efficient energy use without disrupting user routines.

\subsection{Recommender Systems in Commercial and Large Buildings}
\label{sec5:b}
Studies on commercial and large building applications focus on complex environments like offices, shopping centers, and public buildings, which require tailored strategies to balance energy efficiency with occupant comfort and operational demands~\cite{kar2019revicee,mao2019environmental}. Unlike residential buildings, where energy consumption is largely influenced by individual behaviors, commercial and large buildings face challenges due to their scale, diverse occupancy patterns, and significant energy loads. 

Over the past decade, there has been a strong focus on advancing nearly zero energy buildings (nZEBs) that utilize renewable energy sources and sophisticated energy management systems~\cite{cao2016building}. Despite the benefits, the implementation of nZEBs has faced challenges globally, primarily due to the substantial costs involved in their deployment~\cite{huang2018uncertainty}. Consequently, research on sustainable building recommendations has increasingly emphasized improving occupant comfort alongside energy efficiency.

In commercial settings, maintaining occupant comfort while achieving energy efficiency is a primary challenge. Research in this area typically explores the use of advanced sensing and control systems to dynamically adjust building parameters, such as lighting and Heating, Ventilation, and Air Conditioning (HVAC) operations, based on real-time data. Kar \etal~\cite{kar2019revicee} developed a collaborative filtering system that optimizes indoor lighting in open-plan offices by aligning individual comfort with energy efficiency. This system, learning from historical user data gathered via sensors and smart applications, employs machine learning techniques to customize lighting recommendations, achieving an energy consumption reduction of up to 72\% compared to conventional methods. Wei \etal~\cite{wei2018energy} proposed energy-saving strategies by optimizing occupant locations and schedules in commercial buildings. This system offers two types of recommendations: relocating occupants to different spaces and adjusting arrival or departure times to reduce energy use. Using a simulator combined with a Q-learning-based recommendation system, they evaluate the potential energy savings of these adjustments.
Wei \etal~\cite{wei2020deep} developed a system for optimizing energy use in commercial buildings using a deep reinforcement learning model structured as a Markov decision process. This system evaluates building conditions, identifies optimal energy-saving actions, offers recommendations, and dynamically adapts based on user feedback, ensuring scalability and efficiency across extensive building networks. Besides, Pinto \etal~\cite{pinto2018multi} introduced a multi-agent case-based reasoning (CBR) system for optimizing energy management in buildings. This system leverages K-NN clustering to identify patterns from historical data, applies SVM for predictive accuracy, and uses an expert system to fine-tune energy consumption recommendations based on established rules.

General energy management systems in commercial buildings aim to provide holistic solutions that integrate multiple building subsystems to optimize overall energy use. These systems often combine data from various sources, such as sensors, weather forecasts, and occupancy schedules, to develop comprehensive energy-saving strategies. Sardianos \etal~\cite{sardianos2021smart} presented a versatile online recommendation system applicable to both households and large public buildings. By combining sensor data with user habits and feedback, the system generates personalized energy efficiency recommendations tailored to the specific needs of each building type. Siddique \etal~\cite{siddique2023smartbuild} developed a recommendation system that leverages Smart Readiness Indicator (SRI) and Building Information Modeling (BIM) data to enhance building energy efficiency. The system specifically targets improvements in HVAC systems and thermal insulation, offering practical recommendations that significantly reduce energy usage.

Furthermore, there remains a need for recommendation systems that are flexible enough to be applied in both residential and commercial settings. These systems are tailored to meet the unique requirements of various building types, serving as flexible tools for diverse energy management needs. Sardianos \etal~\cite{sardianos2021emergence} proposed a context-aware and explainable recommendation system that generates personalized and persuasive recommendations to optimize energy efficiency. The system accounts for both environmental and economic impacts, making it suitable for various building types, from small residential units to large commercial complexes.

\section{Broader applications of sustainable recommendations}
\label{sec:6}
Beyond the specific applications of sustainable recommender systems in food, travel, and building, these systems have the potential to contribute to sustainability efforts across a broader range of domains. Sustainability-oriented recommendation models are specifically designed to integrate sustainable practices directly into the content of their recommendations, focusing on influencing user behavior across environmental, social, economic, and user-specific dimensions. By actively promoting sustainable behaviors, such as recommending products and services that minimize environmental impacts, encourage energy-saving actions, or promote planned purchasing to reduce waste~\cite{Jnr21,lee2020multi,StarkeWS21}, these models are particularly effective in areas where consumer choices significantly impact environmental outcomes.

This section explores the broader applications of sustainable recommender systems, categorized into four main areas: environmental and ecological sustainability, behavior and social change, economic and productive sustainability, and user-centric sustainable recommendation. This categorization allows us to examine how recommender systems can be specifically designed to address diverse sustainability challenges by minimizing environmental impact, promoting sustainable behaviors, supporting economic activities that align with sustainability goals, and enhancing user engagement in sustainability practices.

\subsection{Environmental and Ecological Sustainability}
\label{sec6:a}
Within the evolving landscape of sustainability-oriented recommendation systems, a specialized segment of models is increasingly dedicated to fostering environmentally responsible behaviors. Leveraging state-of-the-art recommendation algorithms, these models advocate for practices and products that minimize environmental impact, such as reducing pollution and conserving resources, while enhancing biodiversity.

Building upon the foundation of sustainable practices within the realm of recommendation systems, various researchers have advanced methodologies that align technological innovation with ecological preservation. Jnr~\cite{Jnr21} utilized CBR to assist city planners in enhancing smart city initiatives towards sustainability. The CBR recommendation system provides a comprehensive approach to the sustainable development of smart cities, focusing on not only technological and economic development, but also the coordination and integration of society, environment and governance. Lee \etal~\cite{lee2020multi} developed a multi-period product recommender system for the online food market, employing LSTM networks to improve prediction accuracy across multiple purchasing periods. This system encourages planned consumption and efficient purchasing decisions, effectively reducing food waste by minimizing the disposal of perishable items. Onile \etal~\cite{onile2023energy} introduced an optimization method for battery energy storage systems in solar microgrids using multi-agent reinforcement learning, aiming to improve grid stability, enhance consumer comfort, and boost energy efficiency. Besides, Wibowo \etal~\cite{wibowo2017bumblebee} explored biodiversity conservation and eco-friendly recommendation systems, proposing a recommender system designed to promote biodiversity by suggesting bumblebee-friendly plants for domestic gardens.

\subsection{Behavior and Social Change}
\label{sec6:b}
By integrating user-specific data and social influences, sustainable recommendation systems not only optimize personalization but also promote behavioral changes that are critical for environmental sustainability. Through strategies that range from preference elicitation to the incorporation of gamification within social platforms, researchers are uncovering innovative ways to encourage eco-friendly choices, thereby enhancing both user satisfaction and societal shifts towards sustainability.

Sustainable recommendation systems, greatly influenced by social acceptance and behavioral change, primarily focus on altering consumer habits and societal norms towards more sustainable practices~\cite{StarkeWS21}. By employing strategies that encourage eco-friendly behaviors and cultivate a cultural shift towards environmental responsibility, these systems effectively link individual actions to broader social impacts. Knijnenburg \etal~\cite{KnijnenburgWB14} explored how tailored preference elicitation methods in recommendation systems can enhance user satisfaction and encourage energy-saving behaviors. It demonstrates that well-aligned preference elicitation methods significantly improve the adoption of energy-saving measures, thereby increasing user satisfaction with the system. Tomkins \etal~\cite{TomkinsILG18} developed a probabilistic model that enhances the recommendation of sustainable products by integrating freely available domain knowledge, product metadata, and customer purchase patterns. This model improves recommendation accuracy by scoring both products and customers based on sustainability criteria. Silva \etal~\cite{SilvaARFP13} explored the integration of gamification strategies within social networks to enhance user engagement and promote sustainable behaviors, thus contributing to a more sustainable society. This array of studies highlights the diverse and significant ways in which recommendation systems, by focusing on behavior and social change, can be tailored to enhance user experience and actively promote sustainable practices across various sectors.

\subsection{Economic and Productive Sustainability}
\label{sec6:c}
In the realm of recommendation systems, a notable focus is directed towards economic and productive sustainability. This aspect emphasizes promoting economic activities that are both profitable and environmentally responsible, aiming to cultivate a balanced, sustainable economic environment. These innovative systems integrate advanced algorithms to optimize economic performance while adhering to sustainability principles, striving to enhance efficiency, environmental care, and economic viability across various industries.

This section delves into innovative applications of sustainable recommender systems in the industrial field. For example, Capelleveen \etal~\cite{CapelleveenAYZ18} investigated how explicit and implicit knowledge can be integrated in developing a recommendation system for industrial symbiosis, arguing that techniques based on implicit knowledge may offer superior adaptability to the unique challenges of the industrial symbiosis data environment. Some work utilized multi-objective optimization in recommendation systems to advance manufacturing and production industries. These models stand out by integrating diverse objectives that not only target improvements in efficiency and service quality but also align closely with sustainability goals, balancing economic and environmental impacts while catering to the needs of various stakeholders. Liu \etal~\cite{LiuDHLLJ24} developed a recommendation model that enhances cloud manufacturing services by incorporating sustainability and collaboration metrics, optimizing service quality, eco-efficiency, and synergy through multi-objective approaches. Similarly, Pachot \etal~\cite{PachotAAC21} introduced a system designed to help companies diversify and improve their production to meet both economic and environmental goals, also considering the objectives of local authorities and businesses to promote sustainable production systems. Hyunwoo \etal~\cite{hwangbo2019session} developed a session-based recommender system tailored for sustainable digital marketing in the fashion industry, utilizing item session data and attribute session data with feature-weighted algorithms to enhance the accuracy of promoting sustainable fashion products. This collection of studies illustrates the broad application of recommendation systems in promoting economic and productive sustainability across various industries, utilizing techniques ranging from multi-objective optimization to the integration of explicit and implicit knowledge, all designed to enhance service quality, environmental sustainability, and economic efficiency.

\subsection{User-centric Sustainable Recommendation}
\label{sec6:d}
User-centric approaches are increasingly recognized as a crucial component within the evolving realm of sustainable recommendation systems, effectively aligning with individual user preferences to foster lasting sustainability practices. These systems not only cater to immediate user preferences but also adapt over time to sustain engagement and minimize resource wastage. By harnessing advanced algorithms and user feedback, these recommender systems promise to enhance long-term user satisfaction and promote sustainable consumption habits across various platforms. This approach not only aligns with environmental goals but also ensures that recommendations remain relevant and effective, encouraging users to make choices that contribute positively to sustainability.

Besides focusing on specific areas of sustainability, research also explores user-centric approaches. Hyun \etal~\cite{HyunPCY22} introduced the Personalized Interest Sustainability-aware recommender system (PERIS), designed to predict which items users will remain interested in over time, effectively addressing the challenge of evolving user preferences. Lee \etal~\cite{LeeH11} proposed a recommender system that incorporates green marketing principles to steer consumers towards environmentally friendly products. It uses an adaptive fuzzy inference mechanism to evaluate options based on price, features, and green attributes, aiming to boost consumer awareness and preference for green products, thereby supporting environmental sustainability. Sardianos \etal~\cite{SardianosVDAABH20} introduced a context-aware, goal-oriented recommendation system that helps users change their energy consumption habits by prioritizing actions based on user goals and adapting recommendations to user behavior and environmental conditions. In academic collaborations, sustaining partnerships is vital for recommendation systems. Wang \etal~\cite{WangLYKX19} developed the SCORE model to improve academic collaborator recommendations by focusing on the longevity and effectiveness of these partnerships. Using three key sustainability metrics and a random walk with restart algorithm, SCORE prioritizes recommendations that support long-term, successful collaborations.

This subsection delves into how recommender systems can be strategically designed to advance sustainability across diverse domains. This exploration is organized into four distinct categories: environmental and ecological sustainability, which focuses on promoting eco-friendly choices that minimize environmental impact; behavior and social change, which leverages recommendation algorithms to foster sustainable consumer behaviors and societal norms; economic and productive sustainability, which integrates sustainable practices into economic activities by optimizing processes and aligning with sustainability objectives; and user-centric sustainable recommendations, which aim to sustain long-term user engagement with environmentally responsible content to promote ongoing sustainable behaviors.

\section{Sustainable Design of Recommender Models}
\label{sec:7}
In addition to exploring the diverse applications of sustainable recommender systems, it is crucial to consider the sustainability of the models themselves from a computational perspective. The environmental impact of recommender systems is not only determined by the content they recommend but also by the efficiency of the algorithms and computational resources used to generate these recommendations. Recent research has increasingly focused on developing eco-efficient computational models that aim to reduce energy consumption, carbon emissions, and resource usage within recommender systems~\cite{ShrivastavaLMME23,Lu0GZZMTGZ23,STCZMYSMS23}. These models employ energy-efficient algorithms, optimized computational processes, and sustainable hardware configurations to enhance system performance while minimizing environmental impact. This focus on computational sustainability is particularly important in large-scale applications, where small improvements in efficiency can lead to significant reductions in environmental footprint. The subsequent sections will provide an in-depth examination of the strategies and innovations driving the sustainable design of recommender models, highlighting the importance of computational efficiency in achieving broader sustainability goals.

\subsection{Energy Consumption Analysis}
\label{sec7:a}
As recommendation systems advance in complexity, their energy consumption and environmental impact have become pressing concerns. The continuous evolution of large models necessitates significant computational resources, leading to substantial energy use and consequential carbon emissions. This increasing demand highlights the critical need for energy consumption analysis, aiming to balance computational efficiency with sustainability. Understanding the energy footprint of various algorithms is paramount to fostering sustainable computing practices that mitigate the environmental impacts of technological advancements.

\begin{figure}
\centering
\includegraphics[width=0.39\textwidth, trim={3, 4, 2, 2},clip]{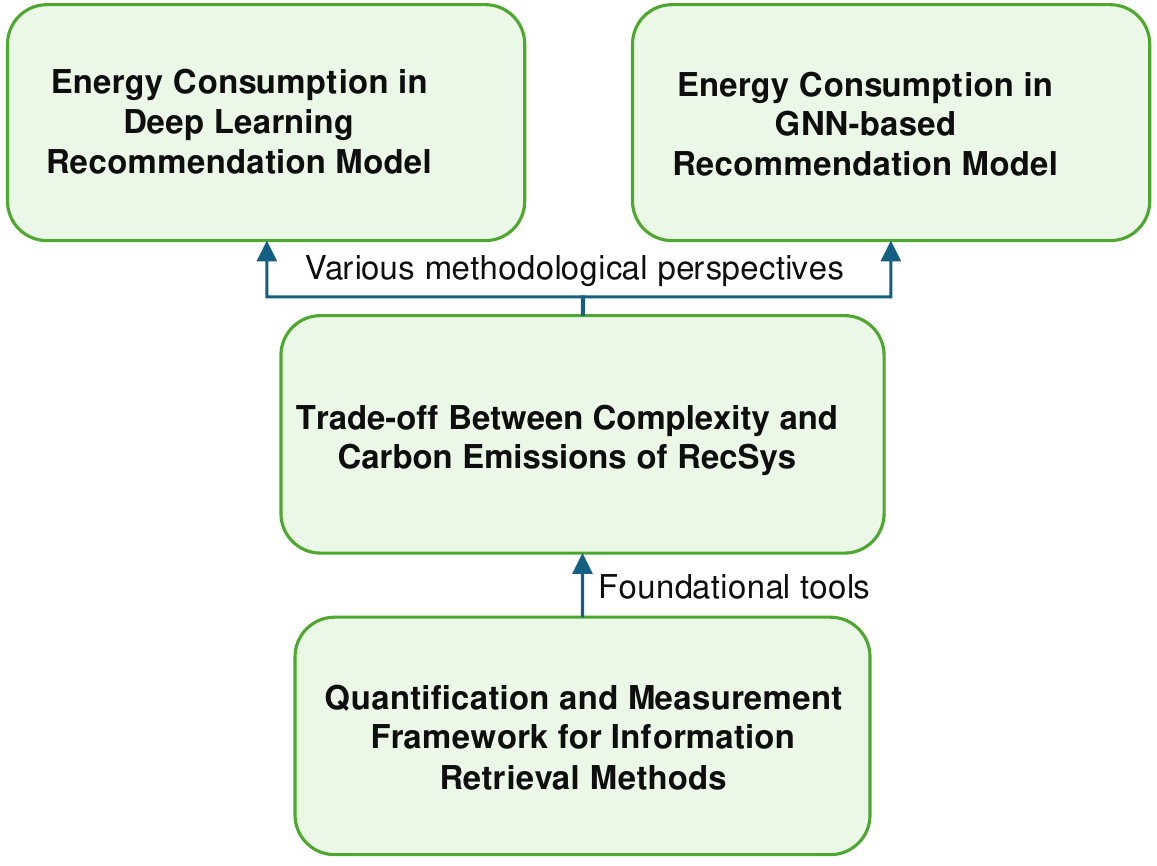}
\caption{Research of energy consumption analysis in RSs.}
\vspace{-12pt}
\label{fig:energy}
\end{figure}

As modern models, including large language models (LLMs), grow increasingly complex and demand vast computational resources, the primary task in advancing sustainability is the analysis of algorithmic energy consumption, with a particular focus on quantifying the specific energy usage of existing research. Training models with billions of parameters, such as GPT-3 or BERT, consumes substantial energy, heavily relying on power-hungry GPUs and large data centers. Strubell \etal~\cite{StrubellGM19}, for instance, found that training large AI models generates carbon emissions comparable to the lifetime emissions of five cars. Similarly, Patterson \etal~\cite{abs-2104-10350} highlighted the massive energy demands and high carbon footprint of training GPT-3. Addressing these energy requirements is not only critical for reducing costs but also essential for mitigating environmental impact, prompting recent research to explore how to reduce the energy consumption of recommendation systems and other AI models, making detailed energy consumption analysis the foundation for achieving more sustainable and responsible computing. Scells \etal~\cite{ScellsZZ22} analyzed the environmental impacts of information retrieval (IR) technologies by quantifying the energy consumption and emissions of the hardware used, and demonstrated the carbon footprint of various IR methods through a series of experiments. Spillo \etal~\cite{SpilloFMMS23} examined the environmental impact of recommendation systems by analyzing the carbon emissions associated with 18 different recommendation algorithms. The study reveals that more complex algorithms often result in minimal performance gains but significantly higher carbon emissions, highlighting the need to balance algorithm efficiency with environmental sustainability. Vente \etal~\cite{vente2024clicks} conducted a comparative analysis of the environmental impact of recommender systems by reviewing 79 experiments from ACM RecSys conferences. This study compared traditional artificial intelligence methods with contemporary deep learning techniques, uncovering that, on average, deep learning approaches produce 42 times more CO$_2$ equivalents. This significant increase in carbon emissions highlights the environmental costs associated with advanced algorithms and underscores the urgent need for integrating sustainable practices within the field. Purificato \etal~\cite{purificato2024eco} analyzed the environmental impact of using GNNs in recommender systems, specifically focusing on carbon emissions and energy consumption, aiming to balance recommendation performance with sustainability.

\subsection{Algorithmic Simplification and Optimization}
\label{sec7:b}
To achieve more efficient computation, both academia and industry are now focused on developing simplified and optimized recommendation systems aimed at creating more sustainable solutions. This ongoing research seeks to reduce the computational complexity and energy consumption of recommendation algorithms, with the goal of balancing high performance with environmental sustainability, ultimately contributing to the development of more eco-friendly recommendation systems. 

\begin{figure}
\centering
\includegraphics[width=0.45\textwidth,trim={5, 0, 2, 0},clip]{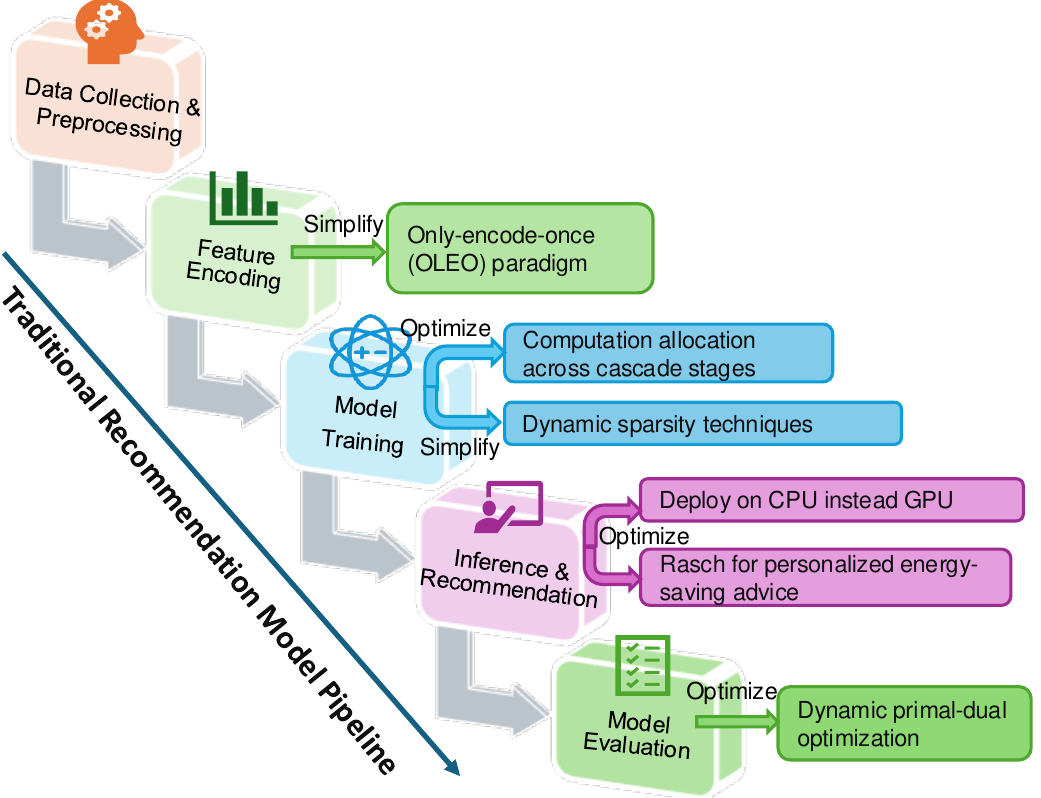}
\caption{Simplification and optimization methods in RSs.}
\vspace{-12pt}
\label{fig:simple}
\end{figure}

Simplifying complex computational processes is an effective approach for sustainable algorithm development. Liu \etal~\cite{LiuZD024} developed GreenRec, a Green AI benchmark for news recommendation systems, which uses an “Only-Encode-Once” training paradigm to pre-train content encoders and cache content vectors, thereby reducing redundant processing and energy consumption. Lu \etal~\cite{Lu0GZZMTGZ23} proposed GreenFlow, a computational allocation framework designed to minimize the carbon footprint and energy demands of recommendation systems. This system dynamically modulates computational complexity using primal-dual optimization techniques, achieving significant reductions in energy use. In a practical application within a mobile industry setting, GreenFlow managed to cut computational needs by 41\%, resulting in a daily energy savings of 5000 kWh and a reduction of 3 tons of carbon emissions, without impacting revenue generation. Yu \etal~\cite{yu2023xsimgcl} simplified graph contrastive learning for efficient recommendations. Shrivastava \etal~\cite{ShrivastavaLMME23} developed the BOLT framework, which is designed to enhance the efficiency of neural recommendation models on CPUs. By utilizing dynamic sparsity, BOLT significantly lowers computational needs, facilitating scalable deployment on more affordable and accessible hardware, promoting sustainability in computational practices. Beyond optimizing and simplifying algorithms, the user interface of recommendation systems is also a key factor in driving energy efficiency and reducing emissions. 
In SelfCF~\cite{SelfCF23} and BM3~\cite{BM3WWW2023}, the authors eliminated the requirement for negative sampling via constractive learning in the recommendation model training process, thereby enhancing computational efficiency.
Other models~\cite{FREEDOM23, DGVAE24} froze the graph structure, thereby preventing it from being modified during the recommendation process. This eliminated the requirement for learning the graph structure during training.
Starke \etal~\cite{StarkeWS17} proposed the Rasch model in a recommender system, which ranks energy-saving measures by their difficulty and aligns them with a user’s ability. The results show that tailored advice reduces user effort, increases system support, and leads to more satisfactory energy-saving actions.

\subsection{Model Compression}
\label{sec7:c}
One effective strategy for developing eco-efficient computational recommendation systems is through model compression techniques. Approaches like pruning~\cite{LiangCNLY23}, quantization~\cite{GZMTLK22,KoYBP0K21}, and knowledge distillation~\cite{WeiTXJH24} are used to simplify recommendation models, reducing their size and complexity while maintaining performance. By cutting down on the computational power needed for training and inference, these methods help to significantly lower energy consumption. 

\begin{figure}
\centering
\includegraphics[width=0.5\textwidth, trim={4, 4, 5, 4},clip]{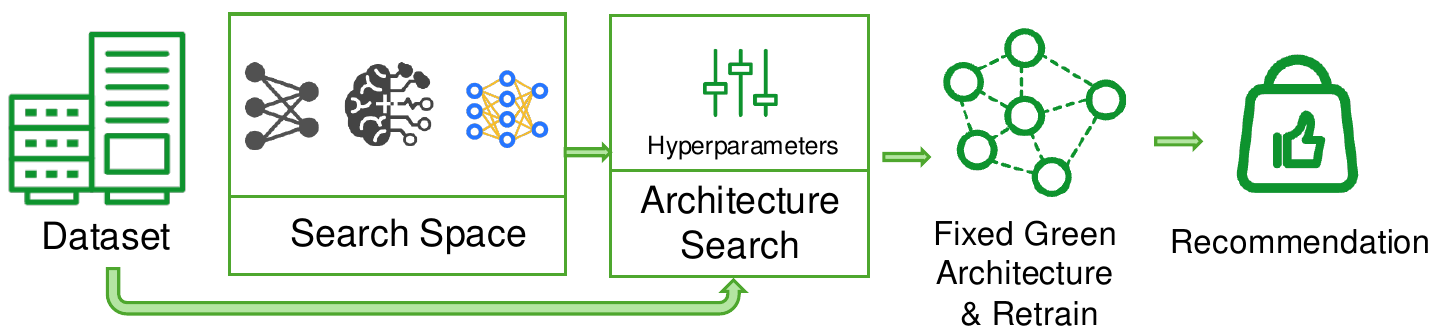}
\caption{Neural architecture search pipeline in recommender systems.}
\vspace{-12pt}
\label{fig:nas}
\end{figure}

Lightweight methodologies are also utilized to develop eco-efficient recommender systems. Zhang \etal~\cite{STCZMYSMS23} introduced SHARK, a model compression approach for large-scale systems that integrates feature selection via first-order Taylor expansion for pruning embedding tables and row-wise quantization. SHARK significantly reduces memory usage and improves query performance without sacrificing accuracy, contributing to eco-efficiency by lowering energy consumption and supporting sustainable, large-scale recommendation systems. Another promising approach for developing sustainable recommender systems is neural architecture search (NAS)~\cite{ZhangCHCDXYL0W23,ChenZHYH22}, which automates the design of neural networks by finding the most efficient architectures, balancing model complexity and performance. These approach help create lightweight, energy-efficient models, supporting the development of sustainable recommender systems by reducing computational demands and energy consumption. For example, Ren \etal~\cite{RenYLLZGZ23} proposed a NAS-based GreenSeq framework designed to optimize the trade-off between computational performance and environmental impact in recommendation models. By creating a multi-layer search space that balances lightweight and heavyweight neural operations, the framework reduces the carbon footprint and energy consumption, particularly during the inference stage of complex models like Transformers.

\subsection{Federated Learning and Edge Computing}
\label{sec7:d}
\begin{figure}
\centering
\includegraphics[width=0.32\textwidth]{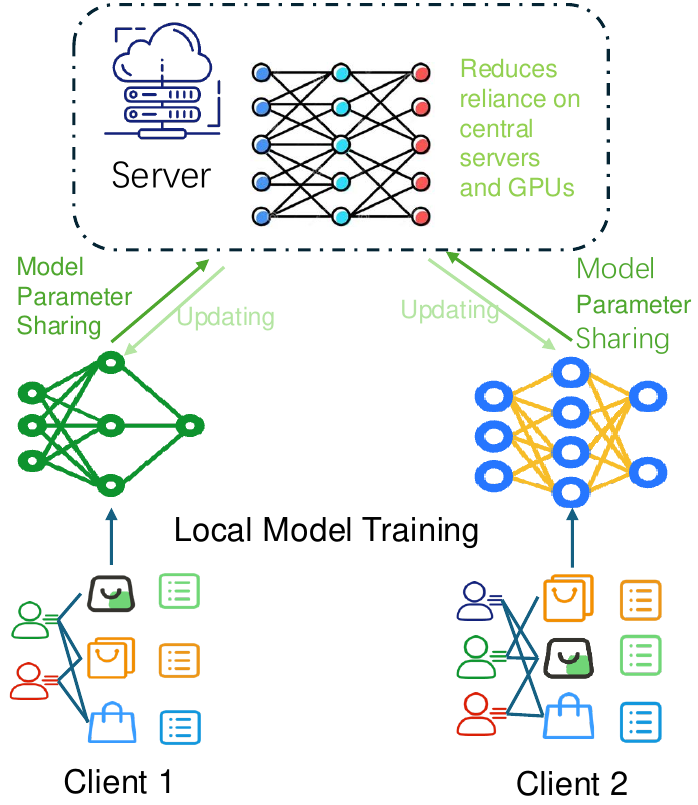}
\caption{A general federated learning model for sustainable recommendation.}
\vspace{-12pt}
\label{fig:fl}
\end{figure}

With the advancement of edge computing and federated learning~\cite{zhang2024transfr,chen2019deep,ding2021graph,zhang2024personalized,zhang2024low,zhang2024communication}, federated recommendation has emerged as a new paradigm in recommendation systems, enabling the offloading of substantial server-side computations to multiple clients~\cite{zhang2024beyond,zhang2024privfr}. These approaches process data locally, reducing reliance on energy-intensive centralized data centers~\cite{SayedHABA22,WangZBYSK24}. By distributing computational tasks and focusing on resource conservation, this strategy supports green objectives, making recommendation systems more sustainable while preserving performance. Maeng \etal~\cite{MaengLMNRW22} highlighted the need to address the interdependence of system and data heterogeneity in federated learning for recommender systems, introducing the RF2 framework to evaluate its impact on fairness and promote more energy-efficient and eco-friendly AI models. Qu \etal~\cite{QuYZCSY24} introduced CDCGNNFed, a novel federated recommendation system that enhances privacy and energy efficiency by allowing users to control data sharing and utilizing localized processing through a cloud-device collaborative graph neural network. This approach balances privacy concerns with recommendation quality while promoting sustainability by reducing energy consumption and the environmental impact of traditional centralized systems. Sayed \etal~\cite{SayedHABA22} introduced a smart, edge-based recommendation system for internet of energy applications that improves energy efficiency by providing real-time, personalized energy-saving advice directly to users’ devices, enhancing privacy and reducing latency through local data processing. Besides, some works focus on POI recommendation, for example, Wang \etal~\cite{wang2020next} proposed a novel on-device POI recommender model (LLRec) on resource-constrained mobile devices, which can be fully compatible with the limited memory space and computing resources. Chen \etal~\cite{chen2018privacy} proposed a distributed POI recommendation method that efficiently distributes computation across multiple edge devices. This approach improves training efficiency and contributes to environmental protection.

The eco-efficient computational approaches discussed emphasize enhancing and optimizing the algorithms themselves, utilizing techniques such as algorithmic simplification and optimization, model compression, and the integration of federated learning and edge computing. These strategies aim to boost computational efficiency, lower energy consumption, and reduce carbon emissions, ultimately fostering the development of more sustainable and eco-friendly recommender systems.

\section{Research Challenges and Future Directions}
\label{sec:8}
{
\subsection{Challenges}
\subsubsection{Public Datasets}
The advancement of sustainable recommender systems is substantially impeded by the paucity of publicly available datasets. Such datasets are indispensable for the rigorous training, testing, and validation of these systems, ensuring the generation of accurate and pertinent recommendations. 
Majority of existing recommendation datasets lack labels that connect directly with sustainability concepts, which are critical for training models to make ecologically responsible recommendations. The limited availability of accessible, diverse, and comprehensive datasets introduces significant methodological constraints in sustainable recommender systems research.
\begin{itemize}[leftmargin=*]
\item {Scientific reproducibility limitations}: Dataset inaccessibility fundamentally compromises the scientific community's capacity to validate and replicate experimental findings, thereby impeding the essential peer verification processes integral to scientific advancement.
\item {Benchmark standardization barriers}: The absence of common datasets precludes the establishment of standardized performance metrics, inhibiting objective comparative analysis of sustainable recommender systems.
\item {Representation biases}: Dataset homogeneity introduces potential systematic biases, potentially yielding models that insufficiently address the diverse requirements of varied user demographics and sustainability contexts, thereby compromising model equity and applicability.
\end{itemize}
Hence, addressing this data deficiency is thus paramount for advancing the field of sustainable recommendation systems and their broader impact on sustainability initiatives.

\subsubsection{Evaluation of Sustainability}
The evaluation of RSs in the context of sustainability presents a complex set of challenges, primarily stemming from the absence of standardized metrics and the heterogeneity of sustainability criteria across diverse domains. 
\begin{itemize}[leftmargin=*]
\item Domain-specific variability: Sustainability metrics vary significantly across different sectors (\eg energy, agriculture, urban planning), making it difficult to establish universally applicable evaluation criteria.
\item Temporal considerations: The long-term impacts of recommendations on sustainability outcomes are often challenging to quantify and predict, necessitating longitudinal studies that are resource-intensive and time-consuming.
\item Multi-objective optimization: Sustainability often involves balancing competing objectives (\eg environmental preservation vs. recommendation accuracy), complicating the development of comprehensive evaluation frameworks.
\end{itemize}

While various ISO standards related to sustainability exist (\eg ISO 14000 series for environmental management, ISO 26000 for social responsibility), translating these broad guidelines into domain-specific, quantifiable metrics for RS evaluation remains a significant challenge. The process of mapping and implementing these standards in specific domains can be hindered by several factors:
\begin{itemize}[leftmargin=*]
\item Abstraction level: ISO standards often provide high-level principles that require substantial interpretation and adaptation for application to RS in specific contexts.
\item Lack of RS-specific guidance: Existing standards are not tailored to the unique characteristics of recommender systems, leaving significant ambiguity in their application.
\item Data availability: The implementation of ISO-aligned metrics frequently requires data that may not be readily available or easily collectable within the RS operational context.
\item Interdisciplinary expertise: Effective mapping of ISO standards to RS evaluation metrics demands collaboration between sustainability experts, domain specialists, and RS researchers—a conjunction that is not always feasible.
\end{itemize}
}
{
\subsection{Directions}
Based on the challenges discussed regarding datasets and evaluation of sustainability in recommender systems, the following research directions are proposed:
\begin{itemize}[leftmargin=*]
\item Development of Sustainability-Centric Datasets:
This direction focuses on creating comprehensive, multi-domain datasets specifically designed for training and evaluating sustainable recommender systems. These datasets should incorporate sustainability indicators (\eg, carbon footprint, social impact scores) alongside traditional recommendation data, enabling researchers to develop and test models that balance user preferences with sustainability goals. This direction supports sustainable practices by providing the foundation for developing recommender systems that inherently consider sustainability factors. By incorporating sustainability metrics into the core dataset, it encourages the creation of algorithms that balance user preferences with environmental and social considerations, potentially leading to more sustainable consumer behaviors and business practices.
\item Long-term Impact Assessment Methodologies:
This direction addresses the challenge of evaluating the long-term sustainability effects of recommendations. It involves developing novel methodologies and tools for predicting and measuring the extended impact of recommender systems on sustainability outcomes. 
This research direction aligns with sustainable practices by focusing on the long-term consequences of recommendations. By developing tools to predict and measure extended impacts, it encourages a more holistic view of sustainability in recommender systems. This approach can lead to the design of systems that optimize for long-term sustainability rather than short-term gains, potentially resulting in more resilient and environmentally responsible recommendation practices.
\end{itemize}
}

\section{Conclusion}
\label{sec:9}
{The urgency of addressing climate change and promoting sustainable practices has become increasingly evident, necessitating technological interventions across multiple domains. Recommender systems, which leverage user behavior patterns to generate personalized suggestions, offer promising capabilities for advancing sustainability initiatives and mitigating climate-related challenges. This survey has comprehensively examined the current progress in leveraging recommender systems for sustainability advancement, encompassing a diverse range of eco-oriented applications and efficient model architectures. The literature reveals promising developments across a broad spectrum of domains including energy conservation, sustainable product recommendations, eco-friendly transportation suggestions, environmentally sensitive building solutions, and among others. However, substantial challenges persist, particularly in the realms of sustainability-focused dataset creation, standardized evaluation metrics, and long-term impact assessment. Future research directions should prioritize the development of multidimensional sustainability metrics frameworks, the creation of domain-specific sustainability datasets, and the formulation of models capable of assessing the long-term environmental and social impacts of recommendations. As the field progresses, interdisciplinary collaboration between computer scientists, sustainability experts, and domain specialists will be crucial in realizing the full potential of recommender systems as a tool for advancing global sustainability goals and combating climate change.}


\bibliographystyle{IEEEtran}
\bibliography{srs_ref}

\vfill

\end{document}